\documentclass[prb,aps,twocolumn,showpacs,preprintnumbers,amsmath,amssymb]{revtex4-1}

\usepackage{graphicx}
\usepackage{dcolumn}
\usepackage[usenames,dvipsnames]{color}

\begin{document}

%useful definitions here
\def\EF{$E_\textrm{F}$}
\def\cred{\color{red}}
\def\cblue{\color{blue}}
\definecolor{dkgreen}{rgb}{0.31,0.49,0.16}
\def\cgreen{\color{dkgreen}}

%\preprint{APS/123-QED}
\title{Strain Tuning of Plasma Frequency in Vanadate, Niobate, and Molybdate Perovskite Oxides}

\author{Arpita Paul}
\affiliation{
Department of Chemical Engineering and Materials Science, University of Minnesota,
Minneapolis, Minnesota 55455, USA
}

\author{Turan Birol}
\affiliation{
Department of Chemical Engineering and Materials Science, University of Minnesota,
Minneapolis, Minnesota 55455, USA
}

\date{\today}

\begin{abstract}
	A novel approach for finding new transparent conductors involves taking advantage of electronic correlations in metallic transition metal oxides, such as SrVO$_3$, to enhance the electronic effective mass and suppress the plasma frequency ($\omega_P$) to infrared. Success of this approach relies on finding a compound with the right electron effective mass and quasiparticle weight $Z$. Biaxial strain can in principle be a fruitful way to manipulate the electronic properties of materials to tune both of these quantities. 
In this study, we elucidate the behavior of the electronic properties of early transition metal oxides SrVO$_3$, SrNbO$_3$, and SrMoO$_3$ under strain, using first principles density functional theory and dynamical mean field theory. We show that strain is not an effective way to manipulate the plasma frequency, but dimensionality of the crystal structure and origin of electronic correlations strongly affect the trends in both $\omega_P$ and $Z$.  
\end{abstract}


\maketitle
%=======================================================================
%%
%%
\section{Introduction}

% Transition metal oxides often have rich electronic and structural phase diagrams.\cite{Khomskii2014_book,Dagotto2005} Electrons in states derived from d orbitals of transtion metal ions interact strongly with both each other and lattice degrees of freedom, and hence are responsible for the plethora of phases observed in these compounds. 
%
Electrons in states derived from d orbitals of transition metal ions in oxides interact strongly with both each other and lattice degrees of freedom, and give rise to rich phase diagrams in these compounds. \cite{Khomskii2014_book,Dagotto2005}
In these phase diagrams, there are often multiple competing phases, and small changes in temperature, boundary conditions (strain, stress, etc.), or chemical composition can lead to significant changes in materials' properties.\cite{Wang2009,Tokura2003,Tokura2000} 
One reason for this sensitivity is the competition between the kinetic and potential energies of electrons, which may result in a regime where neither a delocalized, non-interacting electron picture (commonly used for semiconductors), nor an atomic picture with electrons in local orbitals is applicable. 

Transparent conductors, materials which are good electrical conductors for direct current (DC), and at the same time highly transparent for visible light, are in high demand for applications including, but not limited to, solar cells, touch screens, smart windows, and LED lighting.\cite{Gordon2000, Edwards2004, Kumar2010, Gao2016} 
Most commonly used transparent conductors are metal oxides. One particular oxide, ITO (indium tin oxide), has more than 97\% of the market share of transparent conducting coatings by itself.\cite{Layani2014} However, ITO has various shortcomings, the most important of which is the increasing price of indium.\cite{Kumar2010} Because of this, there is an ongoing search for better transparent conductors. For example, perovskite BaSnO$_3$ has drawn significant recent attention because of its record breaking mobility.\cite{Lee2017} 

A common shortcoming of most transparent conducting oxides is that they rely on doping or alloying a wide bandgap semiconductor. This approach is limited because of possible doping bottlenecks and reduced mobility due to scattering by impurity atoms. An alternative design strategy to find new, superior transparent conductors, put forward by Zhang et al. in 2015,\cite{Zhang2015} is to look for stoichiometric \textit{metals} with a single, energetically isolated, and narrow band crossing the Fermi level.
Metals are typically not transparent because of the plasma reflectivity of the free electron gas.\cite{Fox2001}
In Zhang et al.'s approach, the small bandwidth of the band crossing the Fermi level, in other words the large electron effective mass $m^*$, ensures that the plasma frequency\cite{Fox2001} 
\begin{equation}
	\hbar\omega_p=\sqrt{\frac{ne^2}{\epsilon m^*}}
\end{equation}
is below $\sim 1.6$~eV. This results in high reflectivity only below visible frequencies. (Here, $e$ is the elementary charge, $n$ is the free electron concentration, and $\epsilon$ is the dielectric constant that takes into account the core electrons only.)
Absorption is also suppressed, because the band is energetically isolated from other bands so that there are no interband electronic processes that can absorb visible light. Hence, even though there is strong absorption at high frequencies and essentially 100\% reflectivity in infrared, a metal can have a transparency window in the visible portion of the electromagnetic spectrum.  

This idea can be applied to correlated perovskite oxides such as SrVO$_3$ as well.\cite{Zhang2016} A key observation, put forward in Ref. \onlinecite{Zhang2016} is that while a plasma frequency below $1.6$~eV is necessary, a higher plasma frequency also leads to higher electrical conductivity $\sigma$: 
\begin{equation}
	\sigma=\frac{n e^2 \tau}{m^*}.
\end{equation}
SrVO$_3$ is a mildly correlated metal with $\hbar \omega_p=1.3$~eV, which is close to the upper limit required for transparency, and therefore brings together a high electrical conductivity with the absence of high reflectivity in the visible. While potential of SrVO$_3$ for applications as a transparent conductor is limited because of O--p to V--t$_\textrm{2g}$ interband transitions that lead to a high absorptivity above $\sim 2.5$~eV, it has a high figure of merit.\cite{Zhang2016} It is also important because it proves the potential of correlated transition metal oxides as transparent metals.\cite{Zhang2016} 

\begin{figure}
\includegraphics[width=.6\linewidth]{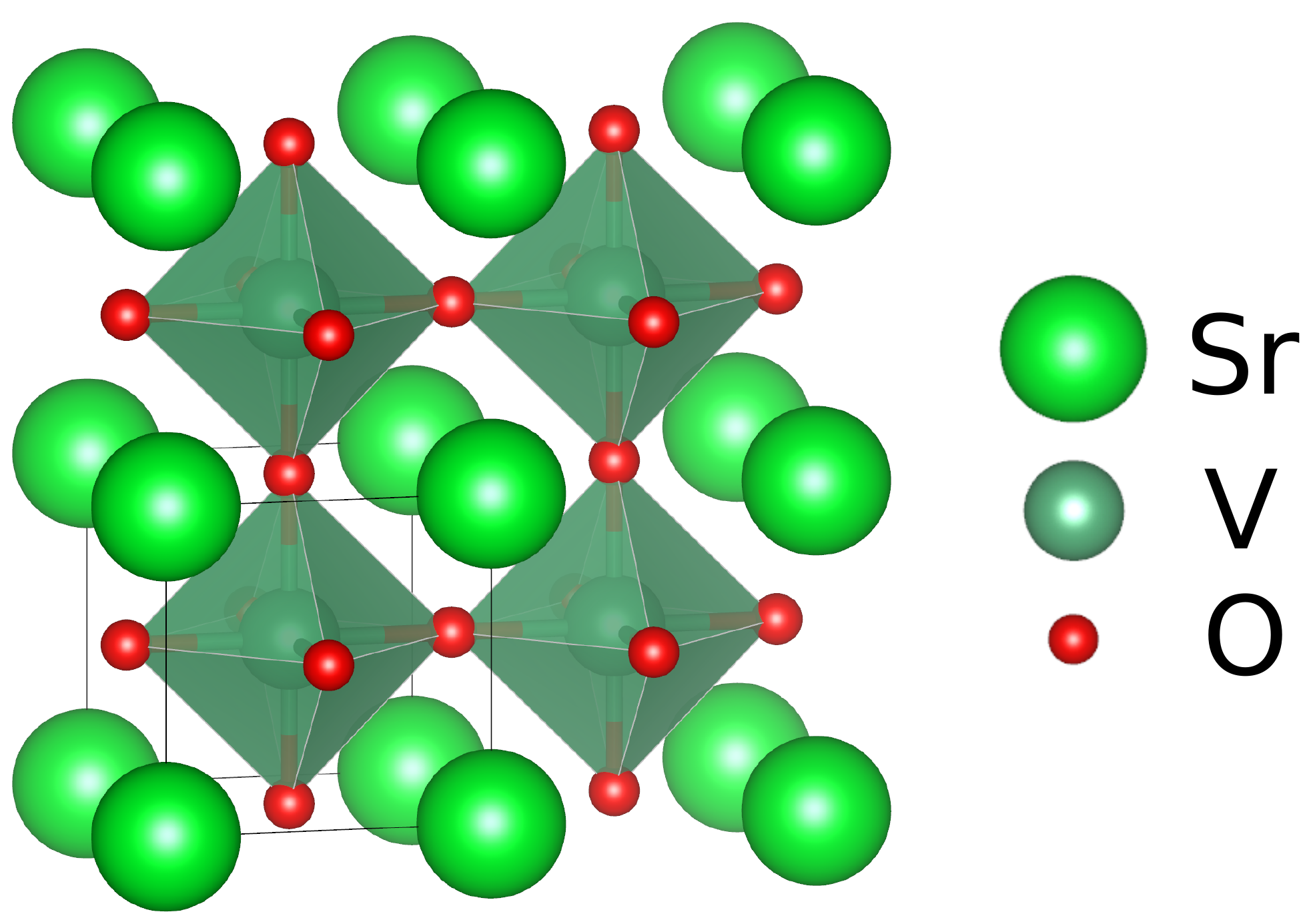}
\caption{(Color Online)
Crystal structure of cubic perovskite SrVO$_3$. V ions are in the center of corner sharing oxygen octahedra.}
\label{fig:SVOCrystal}
\end{figure}

SrVO$_3$ is a metallic compound with free electron concentration $n\sim 10^{22}$, orders of magnitude higher than the average degenerate semiconductors. There are two reasons for its low plasma frequency: \textit{(i)} Due to the corner-sharing geometry of the oxygen octahedra (Fig. \ref{fig:SVOCrystal}) in perovskites, the electron hopping elements between V-d orbitals are much smaller than in typical semiconductors and elemental metals. \textit{(ii)} Also, the electron correlation effects that are often present in 3d transition metal oxides lead to a correlation induced electron effective mass enhancement. In SrVO$_3$, this mass enhancement is by a factor of $Z^{-1}=2$, where $Z$ is the quasiparticle weight. The mass renormalization changes the bandwidth measured by angle-resolved photoemission spectroscopy (ARPES), as well as other quantities such as the Sommerfeld coefficient, from their values calculated from band structure methods such as Density Functional Theory (DFT).\cite{Inoue1998, Yoshida2010} DFT calculations cannot capture the dynamic correlations that give rise to the mass enhancement, and predict the plasma frequency of SrVO$_3$ to be 1.8~eV,\cite{Zhang2016} which is equal to the experimental value only when it is renormalized by $\sqrt{Z}$.

Biaxial strain, imposed by growing thin films on lattice mismatched substrates, is commonly used to alter the structural and electronic properties of transition metal oxides. Drastic changes in ferroelectric, multiferroic, superconducting, and other properties are observed in various systems.\cite{Bozovic2002,Haeni2004,Choi2004,Lee2010_rabe}
While many correlated oxides are known to undergo metal-insulator transitions under strain, to the best of our knowledge strain effects on correlated metals far from a metal-insulator transition is not studied in detail yet. 
Possibilities put forward and questions raised by the recent studies on SrVO$_3$ make understanding how the correlated metals evolve under biaxial strain and heterostructuring a highly relevant problem. 

In this manuscript we study the properties of a series of moderately correlated metallic perovskite oxides (SrVO$_3$, CaVO$_3$, SrNbO$_3$, and SrMoO$_3$) under biaxial strain boundary conditions  by using state of the art first principles methods that are capable of capturing correlation induced mass enhancement (Density Functional Theory + Embedded Dynamical Mean Field Theory (DFT+eDMFT)). Our study differs from other first principles studies which study the effect of biaxial strain in similar materials, such as ref.'s \onlinecite{Sclauzero2016,Beck2018,Dymkowski2014}, in that we do not focus a possible metal-insulator transition, but rather we explore how mass renormalization factor $Z^{-1}$ and plasma frequency $\omega_p$ evolve while the material is still in the metallic region of the phase diagram, far from the metal-insulator transition. 
Our main findings are that \textit{(i)} not surprisingly, correlated metallic properties of SrVO$_3$ do not depend on biaxial strain sensitively, due to the simplicity of the cubic crystal structure, \textit{(ii)} properties of CaVO$_3$ are more strain dependent thanks to oxygen octahedral rotations which couple strongly with both the electronic structure and strain, \textit{(iii)} despite being a 4d transition metal oxide and having a large bandwidth, SrNbO$_3$ has quantitatively observable correlation effects. We also propose that \textit{(iv)} layered transition metal oxide Sr$_2$NbO$_4$, if synthesized, would be a candidate correlated transparent metal that has lower absorptivity than SrVO$_3$. This finding underlines that while the correlated metals far from a metal insulator transition are not sensitive to strain, layering and heterostructuring are promising routes to tune plasma frequencies and correlation strengths of these compounds. \textit{(v)} We conclude by studying the so-called Hund's metal SrMoO$_3$,\cite{Wadati2014} and show that the Hund's coupling induced correlations are more sensitive to strain induced changes in the degeneracy of transition metal orbitals. 

\section{Methods} 

Prediction of crystal structures under biaxial strain boundary conditions are performed using Density Functional Theory and projector augmented plane wave approach as implemented in the Vienna ab initio Simulation Package (VASP).\cite{VASP1, VASP2} Both the out of plane lattice constants, and all the internal ionic coordinates are relaxed to optimize the energy. For CaVO$_3$, where multiple octahedral rotation patterns are possible, the relaxations are initiated with different starting structures to find the pattern that gives the lowest energy. 
Exchange correlation energy is calculated using the PBEsol generalized gradient approximation.\cite{PBEsol} In order to take into account the underestimation of interactions on localized d orbitals, DFT+U approach\cite{DFTU:Dudarev} is used with $U=3.0$~eV for vanadium, and $U=1.5$~eV for niobium and molybdenum d orbitals. (These values are determined by comparing with the experimentally observed cubic lattice constants.) A (shifted) Monkhorst-Pack grid\cite{Monkhorst1976} of $12\times 12\times 12$ k-points for the Brillouin zone of the primitive cell, and equivalent or better grids for supercells are used for crystal structure determination. The energy cutoff for the plane waves is set to 500~eV. Spin orbit coupling is not taken into account in any of the calculations.  

Calculation of the optical properties via DFT is performed using the full-potential (linearized) augmented plane-wave method as implemented in the WIEN2k package.\cite{WIEN2k, Ambroschdraxl2006} Convergence of the plasma frequency requires a particularly fine k-point grid especially in low symmetry crystal structures. We achieved convergence using $\sim 30000$ k-points in the whole Brillouin zone of SrVO$_3$, SrNbO$_3$ and SrMoO$_3$ cubic structures, and $\sim 40000$ k-points in the Brillouin zone of the 4 formula unit supercell of CaVO$_3$. The unscreened plasma frequency, which does not take into account the screening of the core electrons, is calculated from the band structure as\cite{Ambroschdraxl2006} 
\begin{equation}
\omega_{p0,ij}^2=\frac{\hbar^2 e^2}{\pi m^2}\int p_{i, \vec{k}}^{n,n} p_{j, \vec{k}}^{n,n}  \cdot \delta \left(E_{\vec{k}}^n - E_F \right)d\vec{k}  .
\end{equation}
Here, $m$ is the free electron mass, $p_{i, \vec{k}}^{n,m}$ is the $i$'th momentum matrix element at wavevector $\vec{k}$ between bands $n$ and $m$, $E_{\vec{k}}^n$ is the energy of $n$'th band, and $E_F$ is the Fermi level. 
%
%This formula does not take into account the effect of the screening of core electrons, and 
%
The screened and unscreened plasma frequencies are proportional to each other with a relative factor of $\sqrt\epsilon_{\textrm{core}}$, where $\epsilon_{\textrm{core}}$ is the relative permittivity of core electrons.\footnote{In this context, the 'core electrons' include electrons in all the bands that do not cross the Fermi level, which would include core, semicore, and part of the valence electrons in the terminology of first principles methods such as linearized augmented plane-wave methods.} There is no simple implementation to calculate this quantity. Instead, we calculate the screened plasma frequencies from the point where the real part of the dielectric function crosses zero.\cite{Grosso2000, Edwards2004} 
Unless otherwise stated, plasma frequencies we report are screened plasma frequencies.
% which are equal to the frequency where real part of the dielectric function crosses zero.\cite{Grosso2000, Edwards2004}

Fully self consistent DFT+eDMFT calculations are performed using Rutgers DFT+eDMFT package\cite{Haule2010, Haule2018} with the continuous time quantum Monte Carlo impurity solver\cite{Haule2007} and nominal double counting.\cite{Haule2015_dc} A hybridization window of $\mp 10$~eV around the Fermi level is used. Electronic temperature is set to $k_B T = 20$~meV. 
Due to the different levels of screening explicitly taken into account in different methods, DFT+U and DFT+DMFT (and even different implementations of DFT+DMFT) often require different values of Hubbard-$U$. In this study, $U=10$~eV for Vanadium and $U=6$~eV for Niobium and Molybdenum are used along with in the DFT+eDMFT calculations. Hund's coupling is set to $J=0.7$~eV. These $U$ values are shown to be suitable for the particular implementation employed. (See, for example, Ref. \onlinecite{Haule2014}). The coulomb interaction is calculated via the density-density terms only (the so-called Ising approximation). 
While obtaining high quality crystal structures by calculating the forces on atoms using DFT+eDMFT is now technically possible,\cite{Haule2018, Haule2016Forces, Haule2015_energies, Paul2019} due to the extra computational cost this would bring, crystal structures obtained from relaxations using DFT+U are used for the DFT+eDMFT calculations. 

For a system with wavevector independent electronic self energy $\Sigma(\omega)$, the mass renormalization factor is the reciprocal of the quasiparticle weight $Z$, which is obtained from the real part of the electronic self energy obtained from DMFT: 
\begin{equation}
	Z=\frac{1}{1-\left. \frac{\partial \textrm{Re}\Sigma(\omega)}{\partial \omega}\right|_{\omega=0} }.
	\label{equ:Z}
\end{equation}
It is also possible to calculate $Z$ without performing the analytical continuation using the electron self energy on the imaginary axis as performed in, for example, Ref. \onlinecite{Han2016}. In our calculations, we did not see a quantitative disagreement beyond the numerical error between the results that these two methods give. 

In all of our calculations, we followed the standard approach of using \textit{epitaxially strained bulk} boundary conditions to simulate films grown on substrates. This corresponds to performing structural relaxations with two of the three (in-plane) lattice constants fixed, with the third one (out-of plane) optimized along with the atomic coordinates in the unit cell. The monoclinic angle in monoclinic structures is assumed to be close to 90$^\circ$, and hence not optimized. Periodic boundary conditions are imposed along all three unit cell vectors. This approach does not take into account quantum confinement and other finite size effects, and hence our results are valid intrinsically for a film of infinite thickness and no strain relaxation.

\section{$\mathrm{SrVO_3}$}
SrVO$_3$ is one of the rare\cite{Lufaso2001} oxide perovskite compounds that has the undistorted cubic structure (space group Pm$\bar{3}$m) at all temperatures. 
It is metallic, and displays $T^2$ resistivity almost up to room temperature.\cite{Giannakopoulou1995} 
Electronic structure of SrVO$_3$ has been studied using a wide range of theoretical approaches, and it is now a commonly used test bed for correlated electron methods. It was one of the first compounds studied using DFT+DMFT, as well as using GW+DMFT.\cite{Pavarini2004,Sakuma2013,Taranto2013, Werner2016}

\begin{figure}
\includegraphics[width=1.0\linewidth]{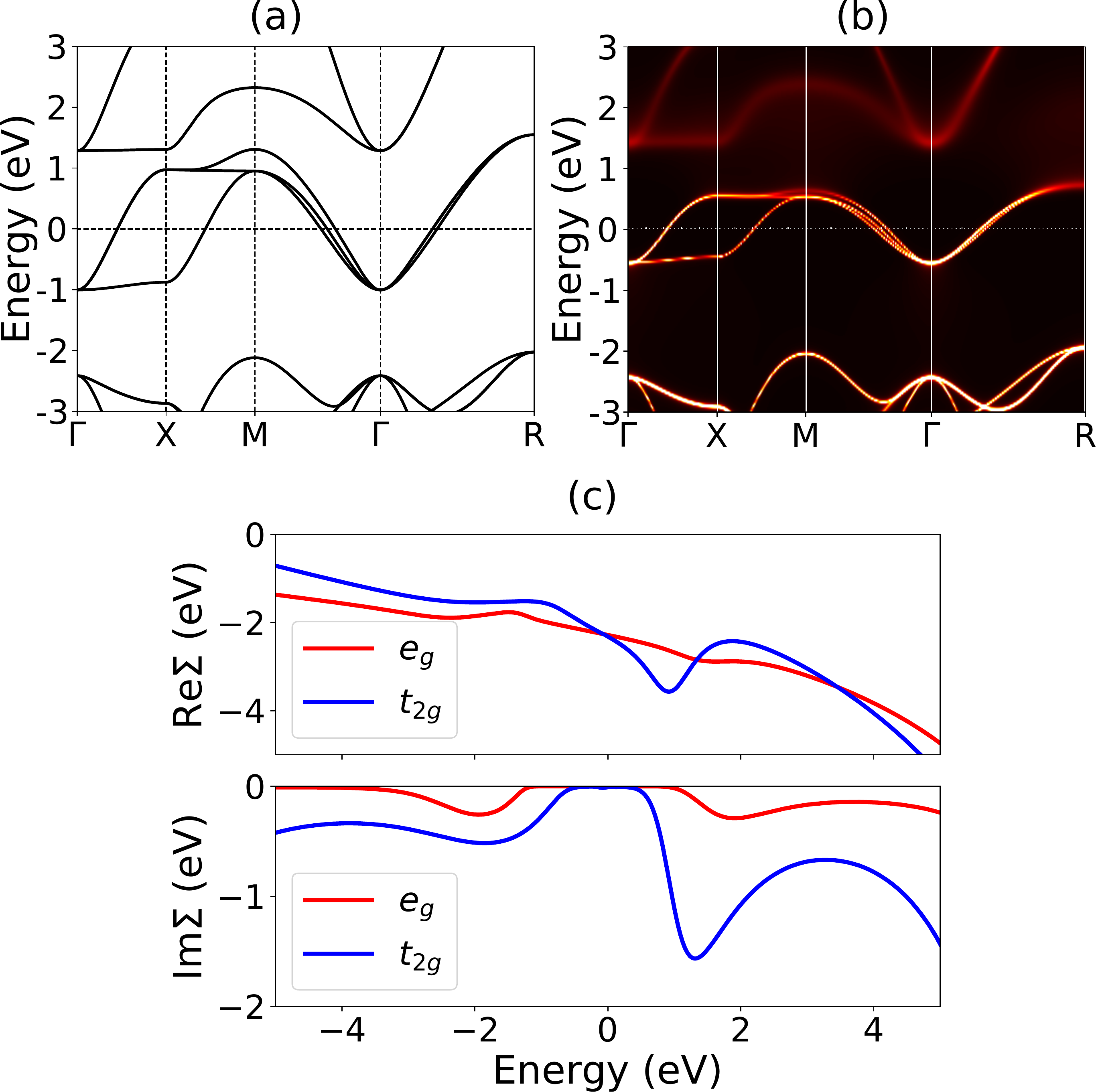}
\caption{(Color Online)
	(a) DFT bandstructure of SrVO$_3$. High symmetry k-point labels refer to those of the simple cubic cell: $\textrm{X}=(\pi/a, 0, 0)$, $\textrm{M}=(\pi/a, \pi/a, 0)$, and $\textrm{R}=(\pi/a, \pi/a, \pi/a)$. (b) DFT+DMFT spectral function of SrVO$_3$, plotted with the same energy scale as panel (a). (c) Real and imaginary parts of DMFT self energy of SrVO$_3$, as a function of energy.}
\label{fig:SVObulk}
\end{figure}

In SrVO$_3$ a single electron occupies the vanadium t$_\textrm{2g}$ bands that cross the Fermi level. These bands are well separated from both the oxygen p and the vanadium e$_\textrm{g}$ bands (Fig. \ref{fig:SVObulk}a). ARPES shows that the bandwidth of the t$_\textrm{2g}$ bands are about half of what DFT predicts.\cite{Takizawa2009, Yoshida2016} This difference has been explained by dynamical electronic correlations on the vanadium site, which give rise to a frequency dependent electronic self energy $\Sigma(\omega)$ that renormalizes the electron effective mass (and hence the bandwidth) as 
\begin{equation}
	m^*=Z^{-1} \cdot m_{\textrm{DFT}}.
\end{equation}
Here, $m_{\textrm{DFT}}$ is the band mass calculated from DFT, and $m^*$ is the effective mass observed in the experiment, and $Z$ is the quasiparticle weight defined in Equation \ref{equ:Z}. 
DFT+DMFT corrects this mass under-estimation by DFT as previously shown many times. The DFT+DMFT spectral function displays 
well defined, quasiparticle-like t$_\textrm{2g}$ bands, as shown in Fig. \ref{fig:SVObulk}b, crossing the Fermi level. 
These bands are about half as wide as they are in DFT calculations, and hence the DFT+DMFT bandwidth 
matches well with the experiment. The mass renormalization factor is $Z\sim 0.55$ for these t$_\textrm{2g}$ bands.  
Real part of the self energy is linear near the Fermi level (Fig. \ref{fig:SVObulk}c). 
Imaginary part, while displaying parabolic behaviour near $E_F$, is very small; ${\textrm{Im}}\Sigma(\omega=0)\sim10$~meV. (We confirmed this value by extrapolating the self energy on the imaginary axis to avoid errors with the maximum entropy analytical continuation process as well. This feature of the self energy is also consistent with that, for example, in Ref. \onlinecite{Sakuma2013}.)
These point to Fermi liquid behavior consistent with the sharp bands in the spectral function. 

\begin{figure}
\includegraphics[width=1.0\linewidth]{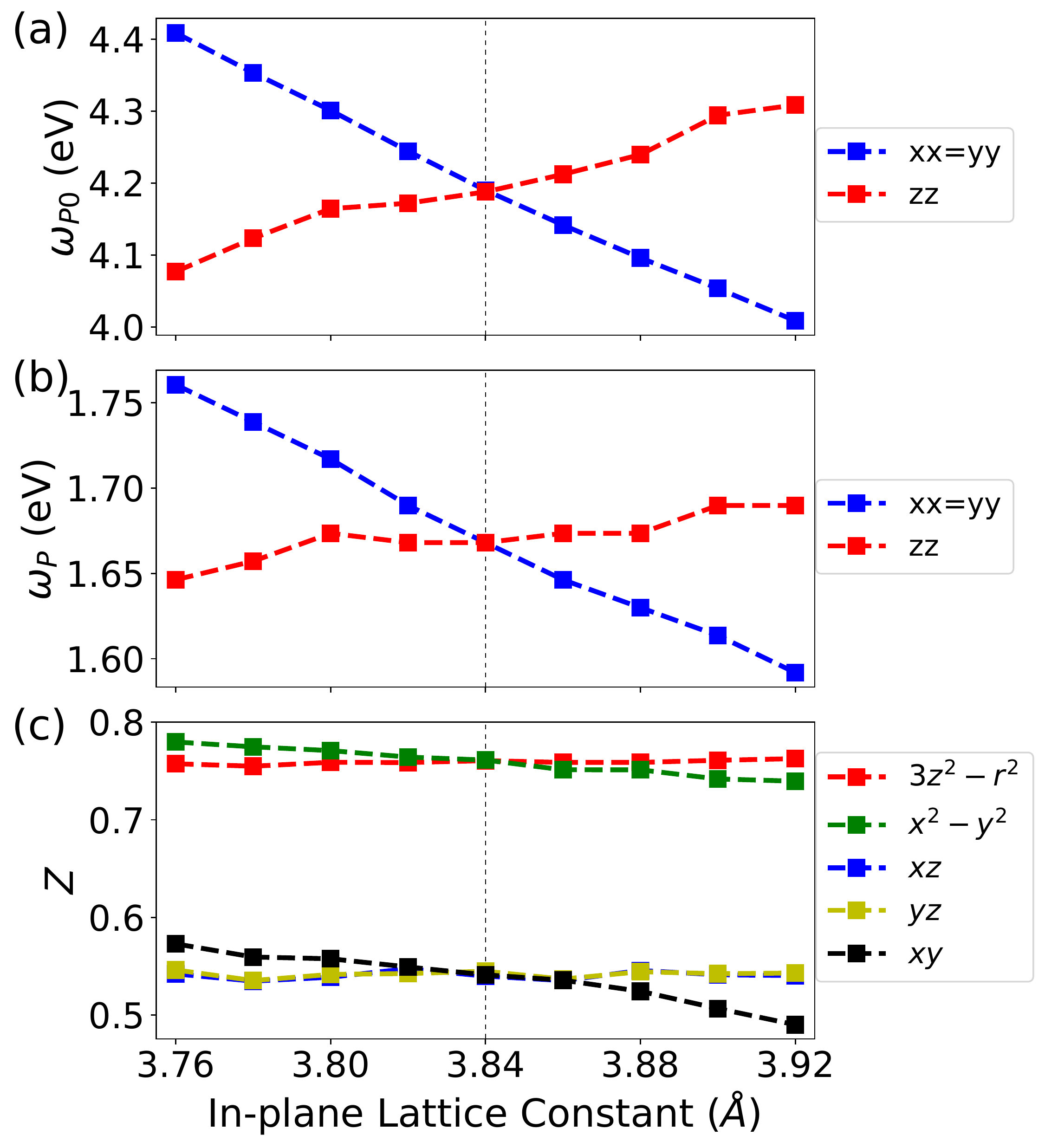}
\caption{(Color Online) Properties of SrVO$_3$ under biaxial strain. (a) Unscreened plasma frequency from DFT. (b) Screened plasma frequency from DFT. (c) Renormalization factor $Z$ from DFT+eDMFT.}
\label{fig:SVOstrain}
\end{figure}

Biaxial strain is known to significantly alter the value of $Z$ in ruthenates,\cite{Burganov2016} and there are examples of strain induced Mott metal-insulator transitions in rare earth titanates.\cite{Dymkowski2014, He2012} In Fig. \ref{fig:SVOstrain}, we present important properties of SrVO$_3$ under strain, obtained from DFT and DFT+DMFT. Fig. \ref{fig:SVOstrain}a is the unscreened plasma frequency calculated from DFT.\cite{Ambroschdraxl2006} This quantity is highly over estimated because it does not take into account the screening by the core electrons. The screened plasma frequency (defined as the point where the real part of the frequency dependent dielectric function crosses zero) is presented in Fig. \ref{fig:SVOstrain}b. Since biaxial strain breaks the cubic symmetry, the plasma frequency depends on the polarization, and has different values for in-plane ($xx=yy$) and out-of-plane ($zz$) polarization. These two components of $\omega_p$ have opposite trends under strain, but neither of them change by more than $\sim 5$\% under a $\sim2$\% strain which is easily achievable in high quality films. 
Similarly, the change in the correlation induced mass renormalization under strain is small: $Z$ is $\sim 0.55$ for all three t$_\textrm{2g}$ orbitals for all strain values considered. The largest change is in $Z$ of the $xy$ orbital, but even that changes by only $\sim10\%$. 

This insensitivity to strain of electronic properties in SrVO$_3$ can be explained by the fact that it is a rather mildly correlated Fermi liquid with low filling (nominally 1 electron in 3 degenerate orbitals), and it has a bandwidth of 2 eV. It is far from the Mott insulator phase boundary, and the strain induced symmetry breaking between its t$_\textrm{2g}$ orbitals is almost negligible compared to the bandwidth. 
Earlier DMFT work by Sclauzero et al.\cite{Sclauzero2016} could obtain a Mott insulating phase in SrVO$_3$, but only under larger values of strain and using a value of $U$ that is larger than the physical value for the particular DMFT implementation used. Our results indicate that room temperature SrVO$_3$ is rather insensitive to strain, and what change its electronic structure exhibits is mostly due to band effects that are reproduced at the DFT level.

\section{$\mathrm{CaVO_3}$}

Alkaline earth A-site cations in oxide perovskites do not contribute to electronic structure around the Fermi level, and as a result, they can be considered as just space filling ions. Going from SrVO$_3$ to CaVO$_3$, there is a significant decrease in unit cell volume (above 6\%) in addition to a reduction in crystal symmetry. Ca is too small for the large oxygen cage of the A-site, and as a result, the oxygen octahedra in CaVO$_3$ are tilted to provide a better A-site coordination environment.\cite{Woodward1997a,Woodward1997b} CaVO$_3$ has the a$^-$a$^-$c$^+$ octahedral rotation pattern in Glazer notation,\cite{Glazer1972} which gives space group Pnma (\#62). 

Effects of the difference in the crystal structures of strontium and calcium vanadates are quite pronounced in ARPES, where a $\sim20\%$ reduction in bandwidth is observed.\cite{Yoshida2010} Early DMFT work of Nekrasov et al.\cite{Nekrasov2005} predicts only a 4\% change in the LDA bandwidth, but a more sizable difference in the mass renormalization near the Fermi level ($Z^{-1}=2.1$ and 2.4 in SrVO$_3$ and CaVO$_3$ respectively). The Sommerfeld coefficient of CaVO$_3$ is measured to be $\sim$10\% higher than that of SrVO$_3$,\cite{Inoue1998} and the optical response of thin films indicate only a small difference in the plasma frequencies.\cite{Zhang2016} 

\begin{figure}
\includegraphics[width=.8\linewidth]{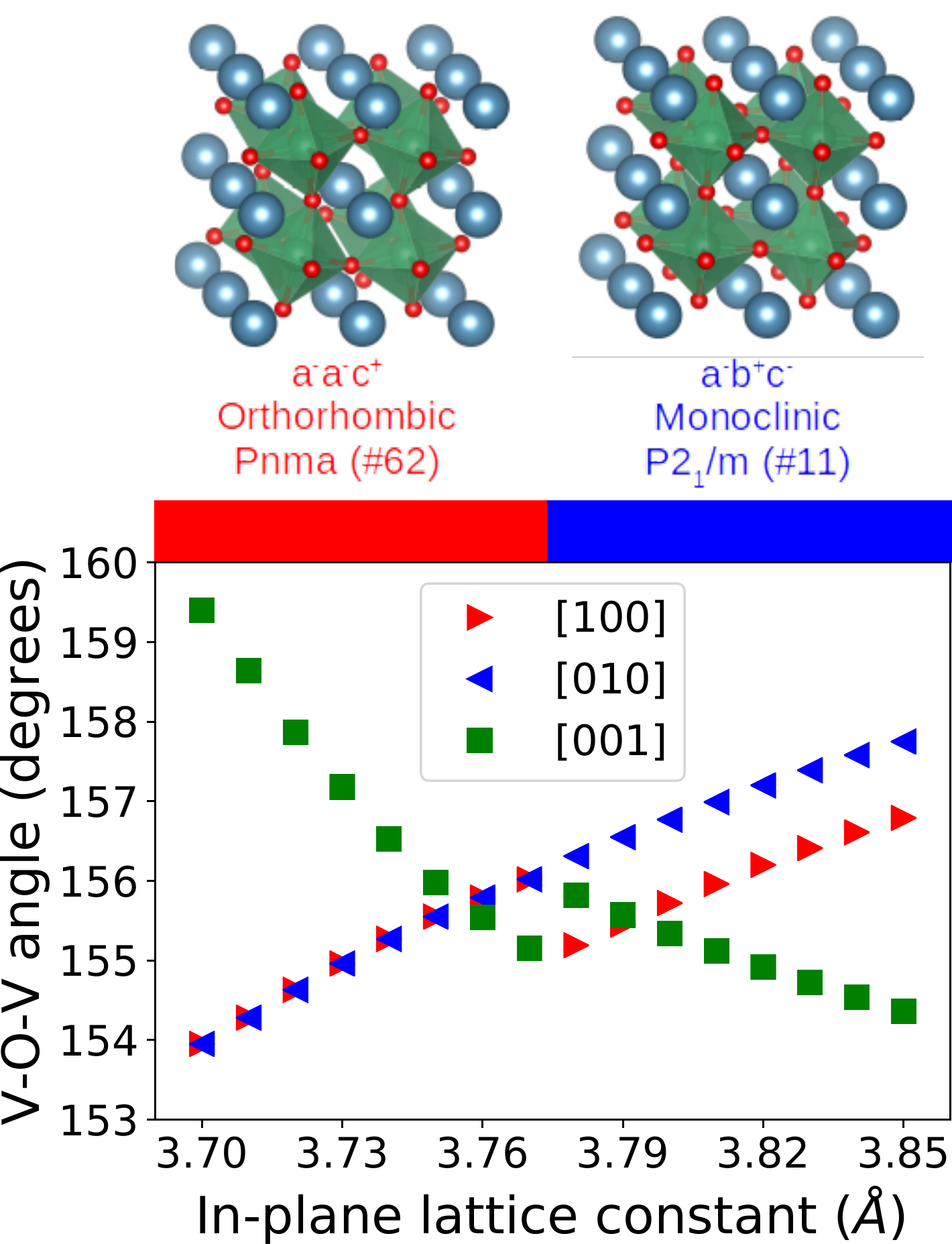}
\caption{(Color Online) V-O-V angles in strained CaVO$_3$ from first principles. Angles along three pseudo-cubic axes are all different under tensile strain due to monoclinic symmetry. }
\label{fig:CVOrot}
\end{figure}

In addition to modifying the electronic structure, octahedral rotations also couple strongly with pressure and strain, and as a result they enhance the effects of biaxial strain on electronic structure. Consequences of this enhanced octahedral rotation mediated strain coupling include, for example, the rich phase diagram of biaxial EuTiO$_3$,\cite{Yang2012_ETO, Birol2013_ETO} change in the magnetic easy axis of SrRuO$_3$,\cite{Lu2015} and the metal insulator transition observed in strained LaTiO$_3$.\cite{Wong2010_LTO, Sclauzero2016} CaVO$_3$ displays large changes in the V-O-V bond angles under strain as well. In Fig. \ref{fig:CVOrot}, we present the V-O-V bond angles of CaVO$_3$ along the three different pseudocubic axes, calculated from first principles. For negative (compressive) strain values, the a$^-$a$^-$c$^+$ rotation pattern is preserved, even though the angles change. In this setting, the $c$ axis of the orthorhombic cell is out of plane, and as a result, biaxial geometry and the presence of the substrate do not reduce the symmetry of the film.  
Like many Pnma perovskites, CaVO$_3$ undergoes a strain induced phase transition near its equilibrium lattice constant (0\% strain) and has a lower symmetry under tensile strain. For positive (tensile) values of strain, the axis around which the rotations are in phase is no longer normal to the film plane. As a result, the rotation pattern becomes a$^+$b$^-$c$^-$, and the symmetry is reduced to monoclinic.\footnote{Part of the reason for this transition is better strain accommodation of the $c$ axis of the Pnma cell, which is of different length than the other two axes. Other examples of the same phenomenon include SrSnO$_3$ and CaTiO$_3$.\cite{Wang2018, Eklund2009}} 

\begin{figure}
\includegraphics[width=1.0\linewidth]{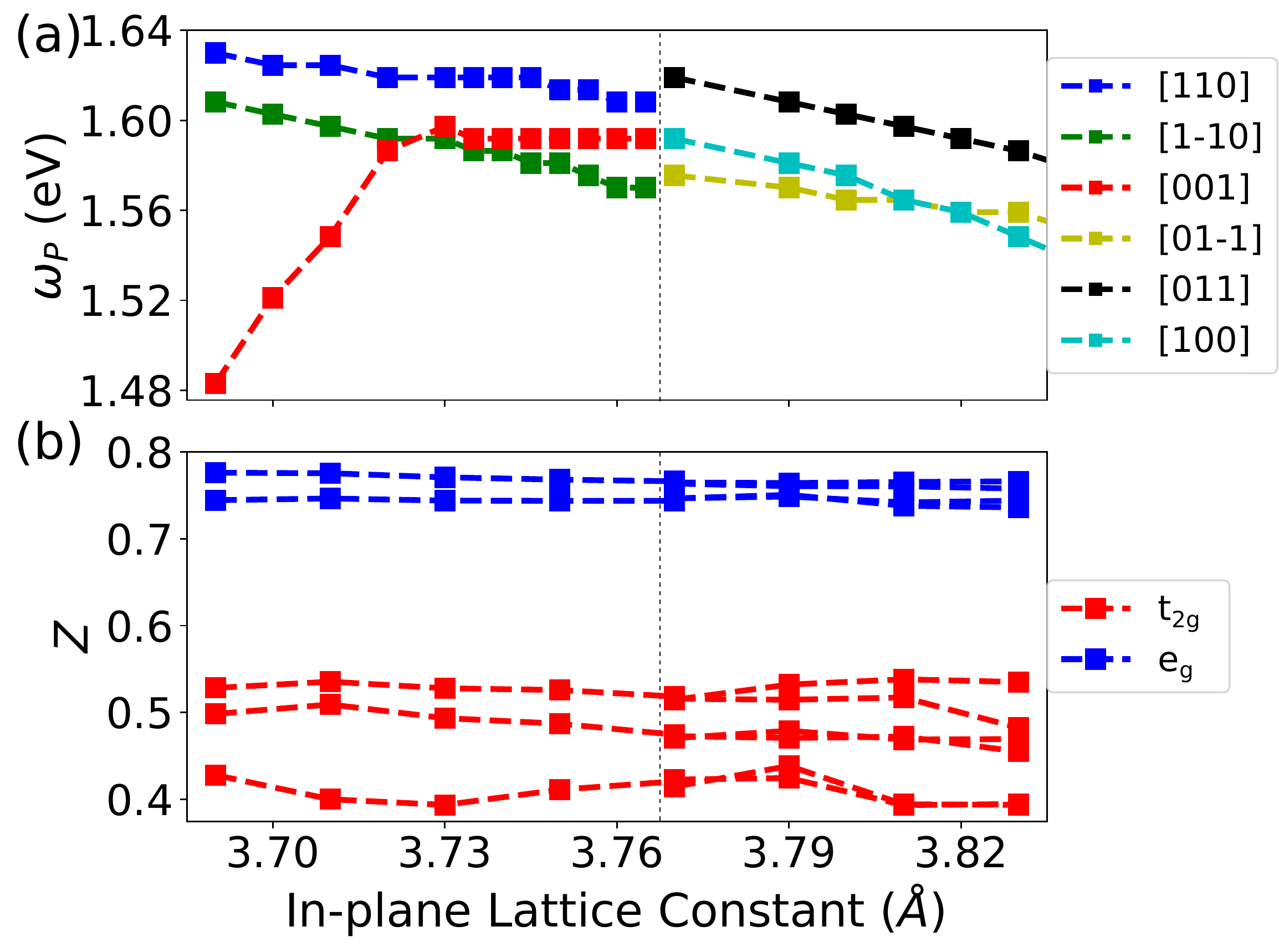}
\caption{(Color Online)
Properties of CaVO$_3$ under biaxial strain. (a) Screened plasma frequency from DFT. Directions given are with respect to the pseudocubic axes. The change in the axes for tensile vs. compressive strain is due to the change in the crystal structure and space group. (b) Renormalization factor $Z$ from DFT+eDMFT. Local vanadium orbitals rotate with the oxygen octahedra, and hence no longer carry simple cubic harmonic characters. Under tensile strain, due to lower symmetry, there are two inequivalent vanadium ions with different $Z$ factors. }
\label{fig:CVOstrain}
\end{figure}

In Fig. \ref{fig:CVOstrain}a, we report the screened plasma frequency of CaVO$_3$ as calculated from DFT. (Note that the choice of the axes for the unit cells used in the calculations are different for the compressive and tensile sides because of the change in the symmetry.) Due to the changing rotation angles with in-plane lattice constant, the change in the plasma frequency at DFT level with strain is larger in CaVO$_3$, compared to SrVO$_3$ (Fig. \ref{fig:SVOstrain}b). On the tensile strain side, all three components of the plasma frequency are suppressed by few percent, comparable to SrVO$_3$. The biggest change is observed in the out of plane polarization direction of the compressively strained films, for which the plasma frequency is reduced by more than 8\% over the course of less than 2\% strain change. 

The $Z$ factor for the t$_\textrm{2g}$ orbitals are split in CaVO$_3$, because the octahedral rotations break the site symmetry of vanadium. (Fig. \ref{fig:CVOstrain}b) We use local coordinate axes for each vanadium ion to define the orbitals initially, and then diagonalize the hybridization function from DFT at zero frequency to define the orbitals that we use for the DMFT calculation. This leads to mixing of t$_\textrm{2g}$ and e$_\textrm{g}$ orbitals (as imposed by the site symmetry). Nevertheless, there are 3 lower-energy orbitals that have t$_\textrm{2g}$-like character and are more correlated, which we refer to as t$_\textrm{2g}$ orbitals for simplicity. Another complication is that the monoclinic spacegroup has two inequivalent vanadium sites which have slightly different $Z$ values. 
Fig. \ref{fig:CVOstrain}b shows that the overall range of values that the value of $Z$ for different orbitals cover as a function of strain is larger in CaVO$_3$ than in SrVO$_3$, however, there is no clear monotonic trend.

Considering both the DFT plasma frequencies and mass enhancement factors from DMFT, the octahedral rotations indeed make the electronic structure of CaVO$_3$ more sensitive to strain. However, this sensitivity does not come along with a simple trend, or large enough changes to be useful for applications where tuning the plasma edge is required. This is in part due to the simple electronic structure of  d$^1$ vanadates. 
Compounds where the orbital degree of freedom is a key factor and which have energy scales that compete with crystal field splittings, such as the so-called Hund's metals,\cite{Yin2011, Georges2013} tend to display more strong strain and octahedral rotation angle dependence of electronic correlations in the metallic phase. 
%, since their electronic correlations are sensitively dependent on the competition 
% between the on-site Hund's coupling and the crystal fields. 
%
We discuss this possibility in more detail in section \ref{sec:SMO}.  

In passing, we note that while our DMFT calculations are performed at finite electronic temperature, the lattice is considered to be at zero temperature as it is the common practice. As a result, the octahedral rotation angles are overestimated in our calculations. This possibly results in a small underestimation of the bandwidth ($W$) in CaVO$_3$, which would result in a smaller $Z$ value as well (since $U/W$ becomes larger). While the $Z$ we calculated for bulk CaVO$_3$ ($Z \sim 0.47$) is between the experimentally measured magnetic susceptibility and specific heat enhancements\cite{Inoue1998}, we predict a larger difference between the mass renormalization factors of SrVO$_3$ to CaVO$_3$, most probably due to this overestimated octahedral rotation angles. ARPES measurements indicate that the bandwidth of CaVO$_3$ is $\sim 20\%$ smaller than that of SrVO$_3$ at room temperature,\cite{Yoshida2010} which is in line with our calculations.

\section{$\mathrm{SrNbO}_3$ and $\mathrm{Sr}_2\mathrm{NbO}_4$}

\begin{figure}
\includegraphics[width=1.0\linewidth]{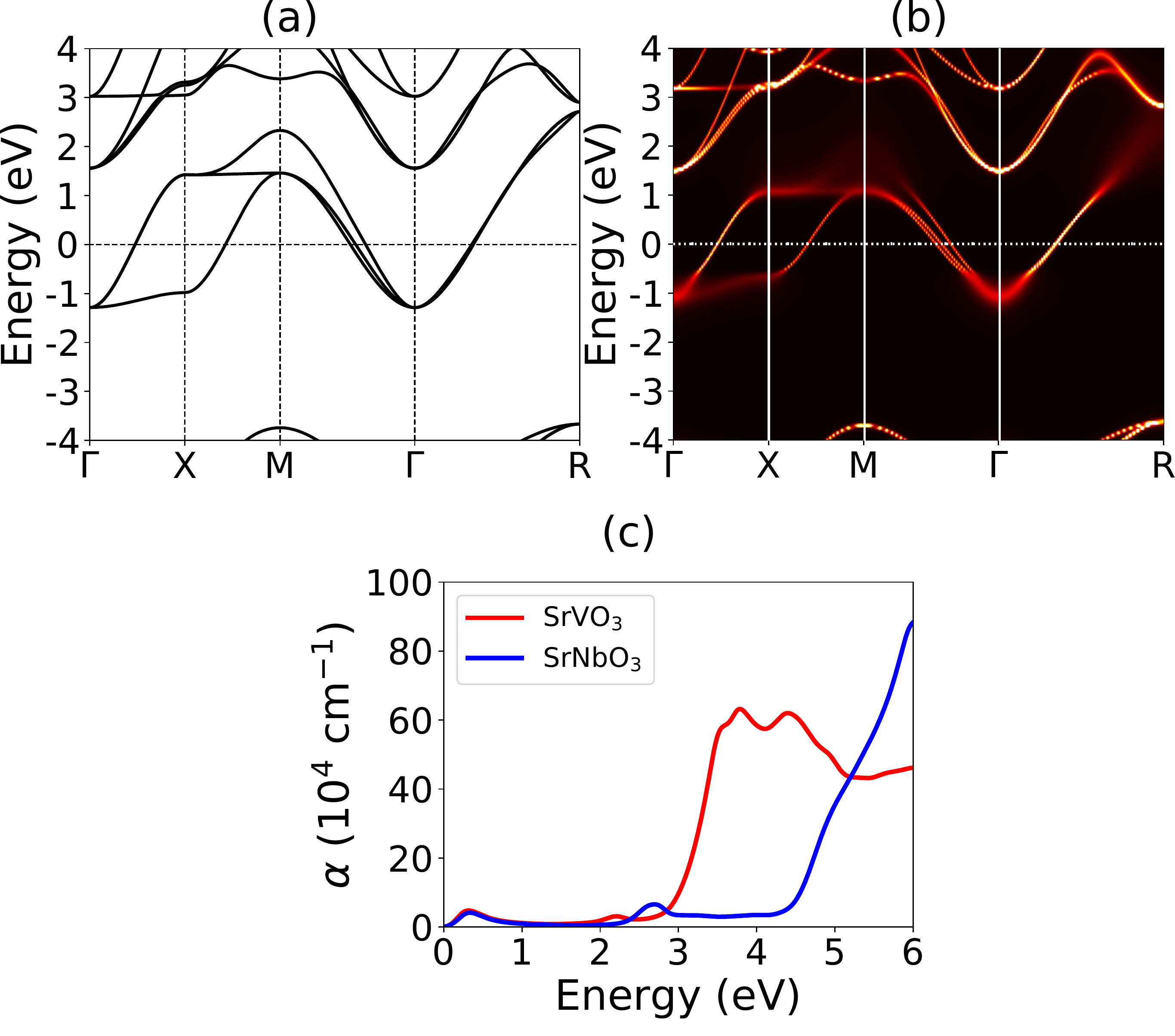}
\caption{(Color Online)
(a) DFT bandstructure of SrNbO$_3$. 
	(b) DFT+DMFT spectral function of SrVO$_3$, plotted with the same energy scale as panel (a)
	(c) Optical absorption coefficients ($\alpha$) of bulk SrVO$_3$ and SrNbO$_3$ from DFT. Both compounds display a small peak due to t$_{\textrm{2g}}$--e$_\textrm{g}$ transitions (at $\sim 2.2$ and $\sim 2.6$ eV for SrVO$_3$ and SrNbO$_3$ respectively) before the onset of high absorptivity at higher energies.}
\label{fig:SNO_bulk}
\end{figure}

SrNbO$_3$ is the 4d analogue of SrVO$_3$. Like SrVO$_3$, it has a single electron on its t$_\textrm{2g}$ bands that cross the Fermi level (Fig. \ref{fig:SNO_bulk}a). Its crystal structure is close to cubic at room temperature with some subtle GdFeO$_3$ type (a$^-$a$^-$c$^+$) octahedral rotations.\cite{Hannerz1999, Macquart2010_SrNbO3, Peng1998} In recent years, there has been an increasing interest in this compound and its optical properties due to its photocatalytic activity.\cite{Xu2012, Efstathiou2013, Wan2017} What makes it also interesting for transparent conductor applications is that the onset of p--d excitations, which results in a sudden upturn of absorbance, is at $\sim 4.5$ eV.\footnote{While there are lower energy Nb--d to Nb--d transitions, they give rise to only a small absorptivity peak around $\sim 2.7$~eV.} This absorption edge is in the ultraviolet, unlike that of SrVO$_3$, which is at $\sim 3$ eV (Fig. \ref{fig:SNO_bulk}c)).  As a result, SrNbO$_3$ has a transparency window that spans part of the visible spectrum.\cite{Wan2017}

SrNbO$_3$ is experimentally observed to have a bright red color when Sr deficient.\cite{Xu2012} The reflectivity of electron deficient Sr$_{1-x}$NbO$_{3+\delta}$ fits well to a Drude model with $\omega_p=1.65$~eV.\cite{Wan2017} First principles calculations which don't take into account the nonstoichiometry are expected to overestimate the plasma frequency. The DFT band structure (Fig. \ref{fig:SNO_bulk}a) gives the plasma frequency of bulk SrNbO$_3$ as $\hbar \omega_p \sim 2.15$ eV, which is consistent with the larger bandwidth of SrNbO$_3$ compared to SrVO$_3$, but is too large compared to the experimentally observed value despite the nonstoichimetry. Electronic correlations effects in this compound, which were claimed to be possible in Ref. \onlinecite{Oka2015}, might explain this large overestimation by DFT. In Fig. \ref{fig:SNO_bulk}b, we present the spectral function of SrNbO$_3$ from DFT+DMFT. 
%
%The spectral function has well defined bands with t$_\textrm{2g}$ character crossing the Fermi level. 
%
$Z$ for the t$_\textrm{2g}$ orbitals is predicted to be $Z=0.72$. This gives $m^*/m_{DFT}\sim 1.2$, so SrNbO$_3$ is a weakly correlated metal. 
This is a surprising observation, because the correlation effects in 4d transition metals oxides with crystal structures that consist of highly connected octahedra (and therefore have a large bandwidth) are usually not expected to be important.
The plasma frequency renormalized from its DFT value is $\hbar\omega_p=1.8$ eV. While this is larger than the experimentally observed $\hbar \omega_p=1.65$~eV, these values are consistent within the experimental error bar in stoichiometry, and the numerical errors in our DFT+DMFT calculations. 

\begin{figure}
\includegraphics[width=1.0\linewidth]{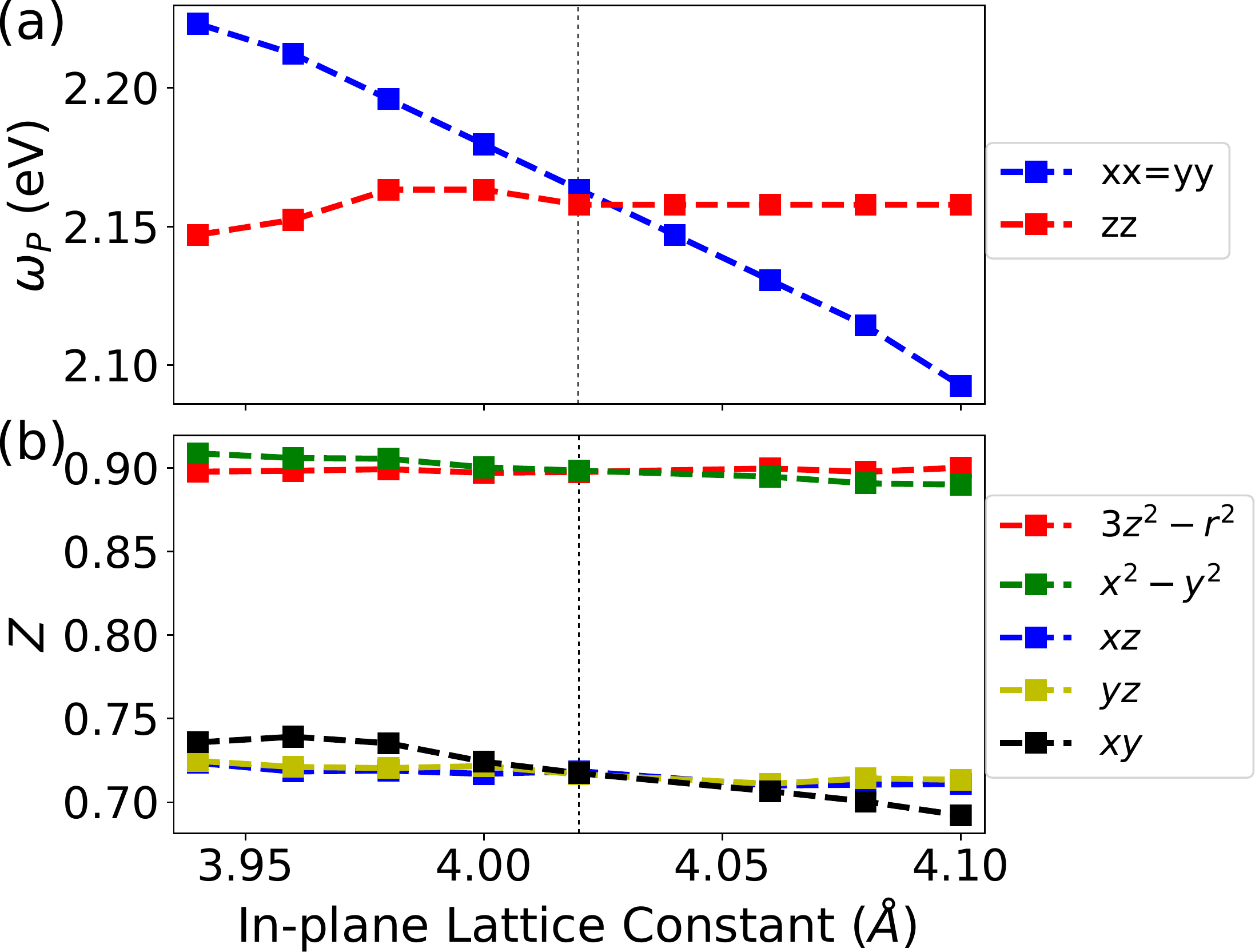}
\caption{(Color Online)
Properties of SrNbO$_3$ under biaxial
strain. (a) Screened plasma frequency from DFT. 
(b) Renormalization factor Z from DFT+eDMFT.}
\label{fig:SNO_strain}
\end{figure}

Even though DMFT underlines the presence of electronic correlation effects in SrNbO$_3$, the strain tunability of these effects are extremely small. 
In Fig. \ref{fig:SNO_strain}, we present the DFT plasma frequency and $Z$ from DMFT for SrNbO$_3$ under strain. Trends in both $\omega_p$ and $Z$ for different orbitals are similar to those observed for SrVO$_3$, but the overall change is smaller. Plasma frequency from DFT changes by $\sim 5\%$ for the $\mp 2\%$ strain range considered, which indicates that the DFT bandwidth is not affected by strain significantly. The change in $Z$ in the same strain range is likewise very small, essentially within the error bars of our DFT+DMFT calculations. This indicates that the effect of strain on the properties of SrNbO$_3$ is negligible. 

Another degree of freedom that can be used as an experimental \textit{knob} to tune and significantly alter the properties of oxide perovskites is layering. For example, the Ruddlesden-Popper structural series, which is a type of layered perovskites, have been used to induce or suppress ferroelectricity, magnetic phases, or superconductivity in various oxides.\cite{Shen1998, Kim2012, Lee2013Birol}
Ruddlesden-Popper phases of SrVO$_3$ have been synthesized.\cite{Nozaki1991} In these layered compounds, reduced bandwidth due to the decreased connectivity of VO$_6$ octahedra result in a very large suppression of conductivity and increased correlation strength: Sr$_2$VO$_3$ is a magnetic Mott insulator,\cite{Nozaki1991,Sugiyama2014, Sakurai2015} and Sr$_3$V$_2$O$_7$ has a resistivity more than an order of magnitude larger than that of SrVO$_3$.\cite{Nozaki1991}
Since SrNbO$_3$ only weakly correlated, Ruddlesden-Popper phases of SrNbO$_3$ are likely to be metallic and they may bring together the lack of optical absorption up to ultraviolet with a plasma frequency that is suppressed to below visible energy range due to a smaller bandwidth. The synthesis of these compounds in the bulk form are challenging, due to both preference of Nb for Nb$^{5+}$ charge state, and the presence of multiple stable phases in the Sr-Nb-O phase diagram (such as SrNbO$_3$, Sr$_2$Nb$_2$O$_7$, etc.\cite{Chan2000}). However, the Ruddlesden-Popper structure is particularly suitable for layer by layer growth of unstable form using advanced synthesis methods such as molecular beam epitaxy.\cite{Haislmaier2016,Haeni2001} To the best of our knowledge, there is only one experimental study of the Ruddlesden-Popper Sr$_2$NbO$_4$ in the last 20 years,\cite{Isawa2001} which reached the seemingly contradictory conclusions that the temperature dependence of resistivity is not metallic, but at room temperature its magnitude is comparable to that of SrNbO$_3$. 

\begin{figure}
\includegraphics[width=.7\linewidth]{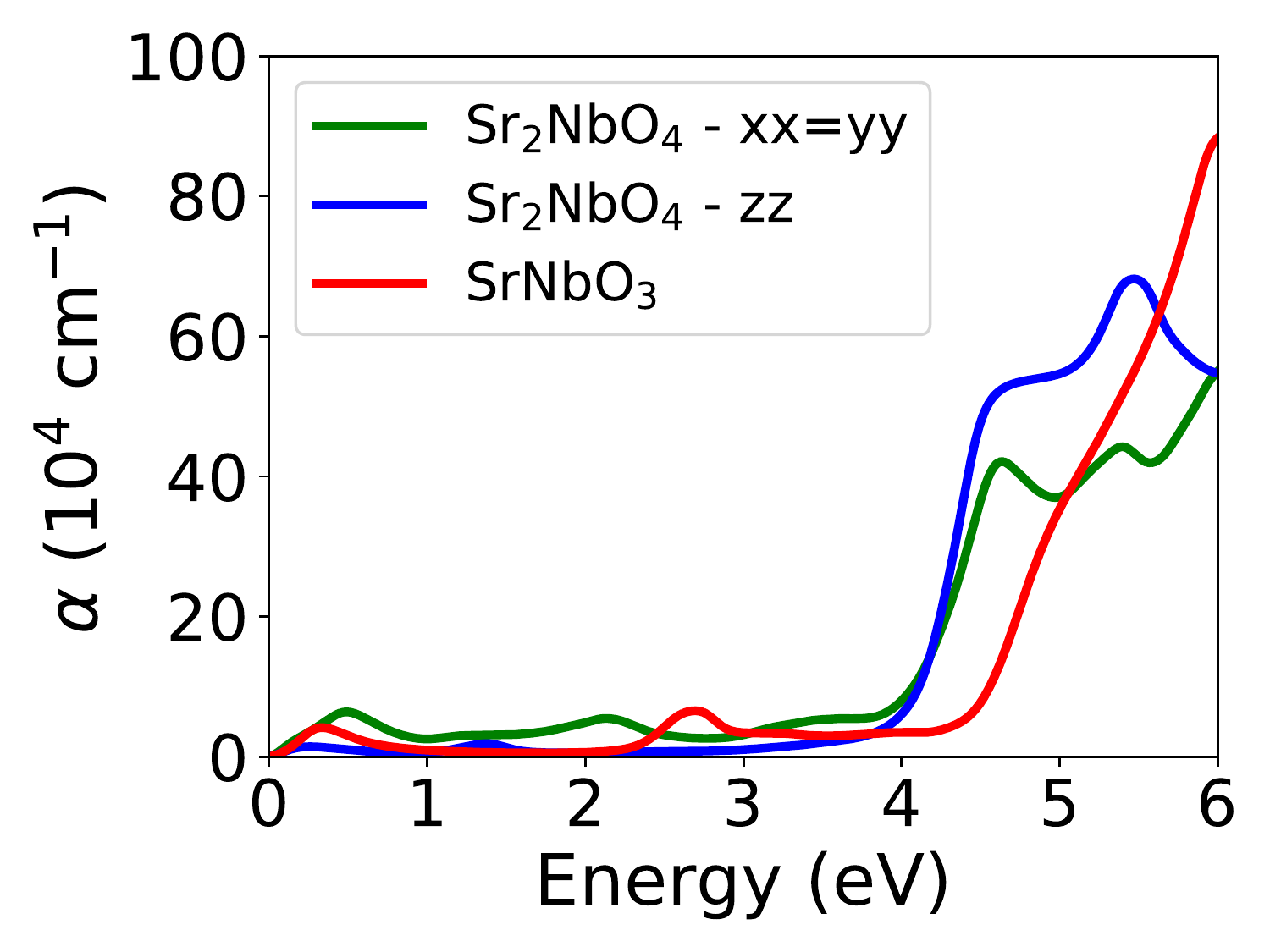}
\caption{(Color Online) Optical absorption coefficient of Sr$_2$NbO$_4$ from DFT for different polarizations of light, compared to that of SrNbO$_3$.}
	\label{fig:SNO214}
\end{figure}

At the DFT level, Sr$_2$NbO$_4$ is metallic, and similar to SrNbO$_3$, it has high absorptivity only in the ultraviolet range (Fig. \ref{fig:SNO214}). 
Its plasma frequency is $\omega_p^{xx=yy}=1.7$ eV and $\omega_p^{zz}=1.0$ eV in the in-plane and out-of-plane (c axis) directions. 
%
%This makes is attractive for possible transparent conductor applications. 
%
% In order to shed light on the electronic properties of Sr$_2$NbO$_4$ and assess its applicability 
% as a transparent conductor, we performed a DFT+DMFT calculation on this layered perovskite. 
%
%At the DFT level, the plasma frequency is suppressed to $\omega_p^{xx=yy}=1.7$ eV and $\omega_p^{zz}=1.0$ eV in the in-plane and out-of-plane (c axis) directions. 
%
DFT+DMFT gives average $Z$ for the t$_\textrm{2g}$ orbitals as $Z\sim0.6$, which renormalizes the in-plane component of the plasma frequency to $1.3$  eV, which is comparable to SrVO$_3$. 
This makes Sr$_2$NbO$_4$ an attractive candidate for applications as a transparent conductor. 
While the strongly anisotropic conductivity is not ideal, other anisotropic materials such as graphene have been implemented as transparent electrodes successfully.\cite{Wu2010}

In summary, we predict Ruddlesden-Popper Sr$_2$NbO$_4$ to be comparable to SrVO$_3$ in terms of electron effective mass on the ab plane, but more transparent for visible light given the high energy onset of absorption in the niobate perovskites. Our calculations demonstrate the feasibility of using crystal structure dimensionality and octahedral connectivity to tune the optical properties of correlated oxides. This can be considered as a new design strategy to predict novel transparent correlated metallic compounds. 

\section{$\mathrm{SrMoO}_3$}
\label{sec:SMO}

\begin{figure}[htbp]
\includegraphics[width=1.0\linewidth]{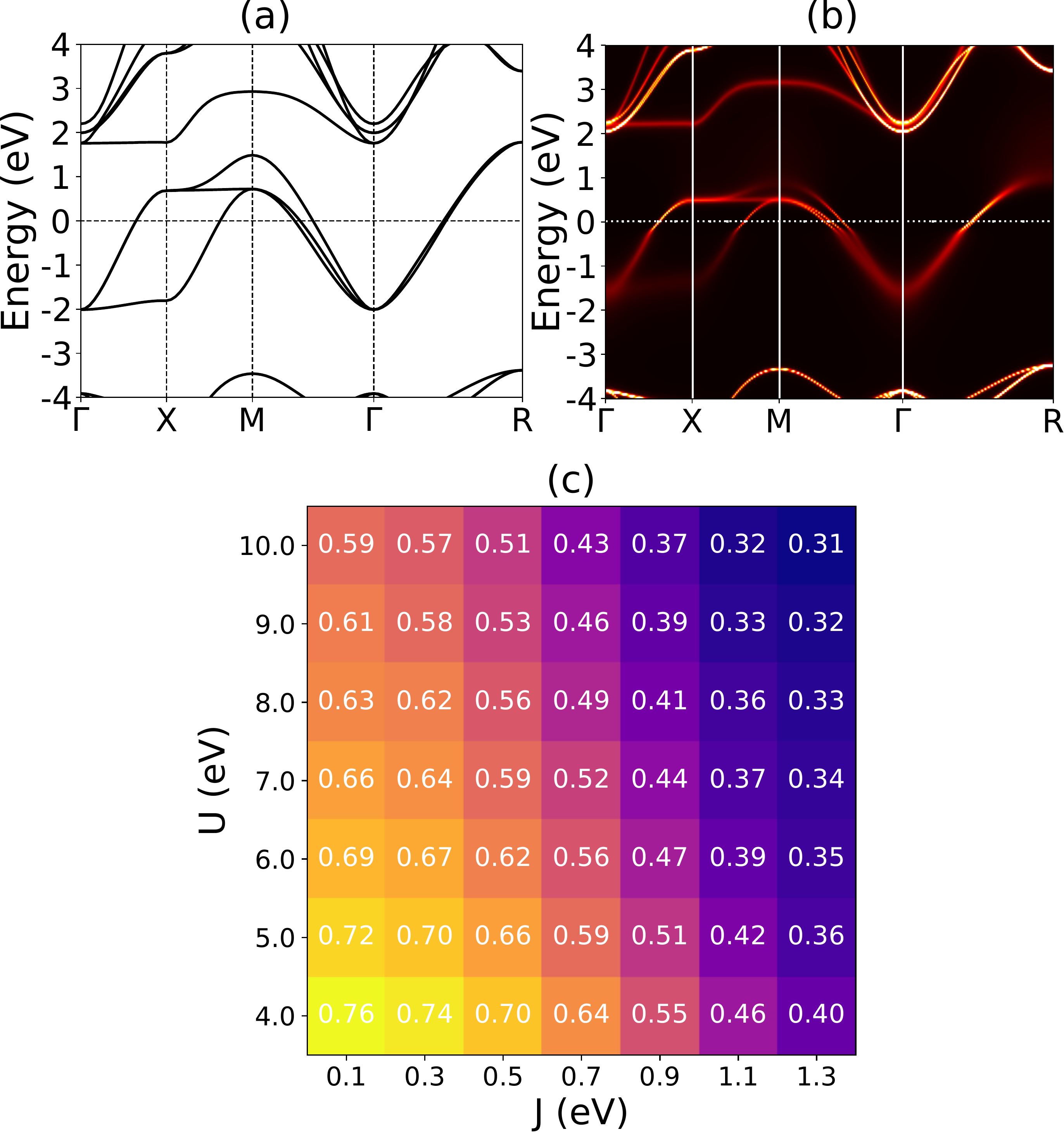}
\caption{(Color Online)
	(a) DFT bandstructure of SrMoO$_3$. Bands crossing the fermi level are of t$_\textrm{2g}$ character, similar to SrVO$_3$ and SrNbO$_3$. (b) DFT+DMFT spectral function of SrMoO$_3$, plotted with the same energy scale as panel (a). The parameters used are $U=6$~eV and $J=0.7$~eV. (c) Quasiparticle weight $Z$ of SrMoO$_3$, calculated from DFT+eDMFT for different values of on-site $U$ and $J$ parameters.}
\label{fig:SMObulk}
\end{figure}

SrMoO$_3$ is another metallic early transition metal oxide with perovskite structure.\cite{Macquart2010_SrMoO3} Optical measurements give its plasma frequency to be $\sim 1.7$~eV.\cite{Mizoguchi2000} What makes this compound particularly interesting is its high conductivity, which broke the record in oxides at room temperature in 2005.\cite{Nagai2005} Mo is a second row transition metal like Nb, and it has the Mo$^{4+}$ charge state with nominally 2 d electrons in SrMoO$_3$. Perovskite SrMoO$_3$ is cubic with no structural distortions above 266~K,\cite{Macquart2010_SrMoO3} and its DFT bandstructure resembles that of SrVO$_3$ and SrNbO$_3$ as expected (Fig. \ref{fig:SMObulk}a). 
Earlier DMFT calculations\cite{Wadati2014} show very little bandwidth renormalization. In Fig. \ref{fig:SMObulk}b, we present the DFT+DMFT spectral density, which reproduce this observation. What makes the electronic structure of SrMoO$_3$ different is that despite the small renormalization of its bandwidth, DFT underestimates its experimentally measured Sommerfeld coefficient by a factor of 2, and its Kadowaki -- Woods ratio is measured to be close to that of heavy fermion compounds.\cite{Nagai2005, Wadati2014} 
The reason of this discrepancy is the driving force of correlations: SrMoO$_3$ is an example of the so-called Hund's metals, where the Hund's J, rather than Hubbard-$U$, is responsible of the bulk of the electronic correlations.\cite{Haule2009, Deng2017, Yin2011,Georges2013} The real part of the electronic self energy of SrMoO$_3$ is not linear in the energy range spanned by the t$_\textrm{2g}$ bands, and hence the value of $Z$ defined by Eq. \ref{equ:Z} at the Fermi level cannot be used to calculate the bandwidth. 
To provide an indirect test of Hund's metallicity in SrMoO$_3$, we calculated $Z$ using different values of $U$ and $J$ parameters in DMFT. Our results, shown in Fig. \ref{fig:SMObulk}c show that the value of $Z$ sensitively depends on the value of $J$ used in the DMFT calculation, but is less sensitive to the value of $U$, as observed in other Hund's metals.\cite{DeMedici2011}  

\begin{figure}[hbtp]
\includegraphics[width=.9\linewidth]{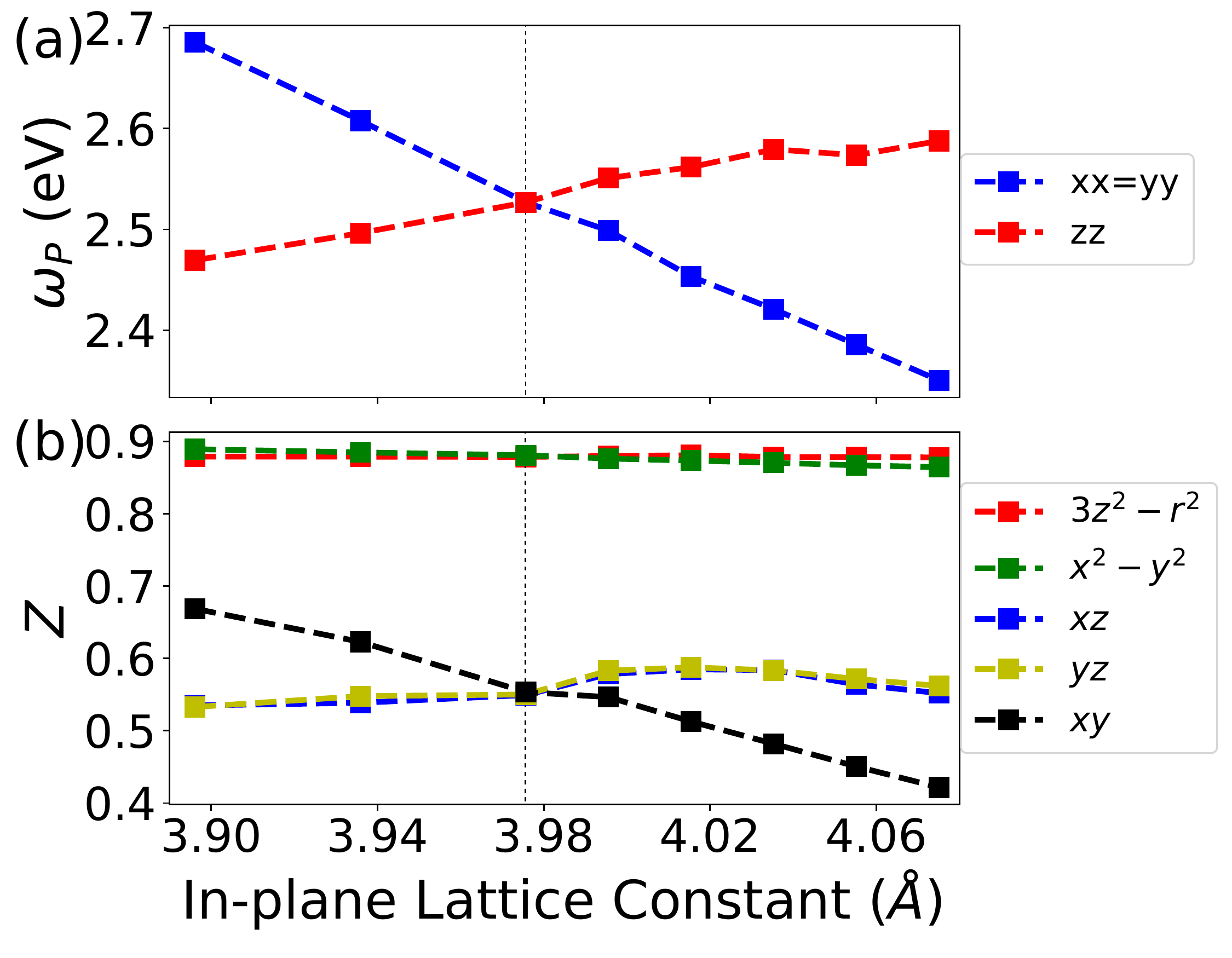}
\caption{(Color Online)
Properties of SrMoO$_3$ under biaxial strain. (a) Screened plasma frequency from DFT. (b) Renormalization factor $Z$ from DFT+eDMFT.}
\label{fig:SMOstrain}
\end{figure}

Correlations induced by Hund's coupling are closely related with the multi-band nature of Hund's metals, and crystal field splittings and differences in widths of different bands have significant effects on the resulting correlation strength and $Z$ parameters. This can result in a stronger biaxial strain effect on the electronic structure, including the plasma frequency and $Z$. In Fig. \ref{fig:SMOstrain}, we present the DFT plasma frequency and orbital dependent $Z$ from DFT+DMFT.
DFT predicts the screened plasma frequency as $\hbar \omega_p \sim 2.5$~eV, which is $\sim 1.9$~eV when scaled by $\sqrt{Z}$. This is larger than the experimentally observed plasma edge at 1.7~eV by an amount that
might be within the combined error bars of the experiment and our calculations. But the simple renormalization of the plasma frequency by $\sqrt{Z}$ is not valid for Hund's metals where the real part of the self energy is not linear, and therefore does not strictly apply to SrMoO$_3$. 
The change in the DFT plasma frequency under strain is similar to that in the other cubic perovskites we considered. On the other hand, the change in the correlation strength, as measured by $Z$, is more pronounced: For the $xy$ orbital, $Z$ goes from $\sim 0.4$ to $0.7$ in the $\mp 2\%$ strain range considered. This is much larger than the $\lesssim 10\%$ change of the same quantity in SrVO$_3$ and SrNbO$_3$, and thus confirms the expectation that biaxial strain effects are more pronounced in Hund's metals than other similar correlated metals. However, the large change is specific only to the $xy$ orbital, and so biaxial strain is not a useful route to tune the plasma frequency of SrMoO$_3$ for transparent conductor applications. 

\section{Conclusions}
% Correlated metallic oxides such as SrVO$_3$ have been in the focus of recent attention because of their 
% plasma frequency that is in the right range where it may provide the optimum combination of carrier 
% concentration and effective mass. 
We performed a series of DFT+DMFT calculations on SrVO$_3$, CaVO$_3$, SrNbO$_3$, and SrMoO$_3$ to 
assess the applicability of biaxial strain as a means to tune the plasma frequency of these compounds. 
These materials display weak to moderate correlation strength, as measured by their $Z$ values in the $\sim 0.50 - 0.75$ range. 
Our calculations show that neither the plasma frequency at the DFT level, nor the correlation induced mass enhancement $Z$ depend
very sensitively on strain in these compounds. 
The presence of octahedral rotations, such as those in CaVO$_3$, or dominant role of Hund's coupling in driving the correlations, such as in SrMoO$_3$, make the strain dependence of $\omega_p$ and $Z$ stronger. However, this stronger trends are direction and orbital dependent, and an overall suppression of plasma frequency for all polarization components is not observed.

A noteworthy result of our calculations is that SrNbO$_3$, which has high orbital degeneracy and $\sim3$ eV wide t$_\textrm{2g}$ bands, has 
nonnegligable electronic correlations as seen from its suppressed plasma frequency. This supports the observations of Ref. \onlinecite{Oka2015}. While the plasma frequency is not suppressed enough to make SrNbO$_3$ optically transparent, Ruddlesden-Popper Sr$_2$NbO$_4$ is more promising: Decreased octahedral connectivity in the Ruddlesden-Popper structure results in both narrower bands in DFT and a smaller $Z$ in DMFT. The plasma frequency is suppressed below the visible range in this compound.  We propose strategies that exploit layering of crystal structures, rather than strain, as a more promising materials design approach to tune the plasma frequency of materials for transparent conductor applications.

\begin{acknowledgements}
This work was supported by NSF DMREF Grant DMR-1629260. We acknowledge the Minnesota Supercomputing Institute (MSI) at the University of Minnesota for providing resources that contributed to the research results reported within this paper. 
\end{acknowledgements}


\begin{thebibliography}{90}%
\makeatletter
\providecommand \@ifxundefined [1]{%
 \@ifx{#1\undefined}
}%
\providecommand \@ifnum [1]{%
 \ifnum #1\expandafter \@firstoftwo
 \else \expandafter \@secondoftwo
 \fi
}%
\providecommand \@ifx [1]{%
 \ifx #1\expandafter \@firstoftwo
 \else \expandafter \@secondoftwo
 \fi
}%
\providecommand \natexlab [1]{#1}%
\providecommand \enquote  [1]{``#1''}%
\providecommand \bibnamefont  [1]{#1}%
\providecommand \bibfnamefont [1]{#1}%
\providecommand \citenamefont [1]{#1}%
\providecommand \href@noop [0]{\@secondoftwo}%
\providecommand \href [0]{\begingroup \@sanitize@url \@href}%
\providecommand \@href[1]{\@@startlink{#1}\@@href}%
\providecommand \@@href[1]{\endgroup#1\@@endlink}%
\providecommand \@sanitize@url [0]{\catcode `\\12\catcode `\$12\catcode
  `\&12\catcode `\#12\catcode `\^12\catcode `\_12\catcode `\%12\relax}%
\providecommand \@@startlink[1]{}%
\providecommand \@@endlink[0]{}%
\providecommand \url  [0]{\begingroup\@sanitize@url \@url }%
\providecommand \@url [1]{\endgroup\@href {#1}{\urlprefix }}%
\providecommand \urlprefix  [0]{URL }%
\providecommand \Eprint [0]{\href }%
\providecommand \doibase [0]{http://dx.doi.org/}%
\providecommand \selectlanguage [0]{\@gobble}%
\providecommand \bibinfo  [0]{\@secondoftwo}%
\providecommand \bibfield  [0]{\@secondoftwo}%
\providecommand \translation [1]{[#1]}%
\providecommand \BibitemOpen [0]{}%
\providecommand \bibitemStop [0]{}%
\providecommand \bibitemNoStop [0]{.\EOS\space}%
\providecommand \EOS [0]{\spacefactor3000\relax}%
\providecommand \BibitemShut  [1]{\csname bibitem#1\endcsname}%
\let\auto@bib@innerbib\@empty
%</preamble>
\bibitem [{\citenamefont {Khomskii}(2014)}]{Khomskii2014_book}%
  \BibitemOpen
  \bibfield  {author} {\bibinfo {author} {\bibfnamefont {D.~I.}\ \bibnamefont
  {Khomskii}},\ }\href {https://books.google.com/books?id=NZIZBQAAQBAJ} {\emph
  {\bibinfo {title} {{Transition Metal Compounds}}}}\ (\bibinfo  {publisher}
  {Cambridge University Press},\ \bibinfo {year} {2014})\BibitemShut {NoStop}%
\bibitem [{\citenamefont {Dagotto}(2005)}]{Dagotto2005}%
  \BibitemOpen
  \bibfield  {author} {\bibinfo {author} {\bibfnamefont {E.}~\bibnamefont
  {Dagotto}},\ }\href {\doibase 10.1126/science.1107559} {\bibfield  {journal}
  {\bibinfo  {journal} {Science}\ }\textbf {\bibinfo {volume} {309}},\ \bibinfo
  {pages} {257} (\bibinfo {year} {2005})}\BibitemShut {NoStop}%
\bibitem [{\citenamefont {Wang}\ \emph {et~al.}(2009)\citenamefont {Wang},
  \citenamefont {Liu},\ and\ \citenamefont {Ren}}]{Wang2009}%
  \BibitemOpen
  \bibfield  {author} {\bibinfo {author} {\bibfnamefont {K.~F.}\ \bibnamefont
  {Wang}}, \bibinfo {author} {\bibfnamefont {J.~M.}\ \bibnamefont {Liu}}, \
  and\ \bibinfo {author} {\bibfnamefont {Z.~F.}\ \bibnamefont {Ren}},\ }\href
  {\doibase 10.1080/00018730902920554} {\bibfield  {journal} {\bibinfo
  {journal} {Advances in Physics}\ }\textbf {\bibinfo {volume} {58}},\ \bibinfo
  {pages} {321} (\bibinfo {year} {2009})}\BibitemShut {NoStop}%
\bibitem [{\citenamefont {Tokura}(2003)}]{Tokura2003}%
  \BibitemOpen
  \bibfield  {author} {\bibinfo {author} {\bibfnamefont {Y.}~\bibnamefont
  {Tokura}},\ }\href {\doibase 10.1063/1.1603080} {\bibfield  {journal}
  {\bibinfo  {journal} {Physics Today}\ }\textbf {\bibinfo {volume} {56}},\
  \bibinfo {pages} {50} (\bibinfo {year} {2003})}\BibitemShut {NoStop}%
\bibitem [{\citenamefont {Tokura}(2000)}]{Tokura2000}%
  \BibitemOpen
  \bibfield  {author} {\bibinfo {author} {\bibfnamefont {Y.}~\bibnamefont
  {Tokura}},\ }\href {\doibase 10.1126/science.288.5465.462} {\bibfield
  {journal} {\bibinfo  {journal} {Science}\ }\textbf {\bibinfo {volume}
  {288}},\ \bibinfo {pages} {462} (\bibinfo {year} {2000})}\BibitemShut
  {NoStop}%
\bibitem [{\citenamefont {Gordon}(2000)}]{Gordon2000}%
  \BibitemOpen
  \bibfield  {author} {\bibinfo {author} {\bibfnamefont {R.~G.}\ \bibnamefont
  {Gordon}},\ }\href {\doibase 10.1557/mrs2000.151} {\bibfield  {journal}
  {\bibinfo  {journal} {MRS Bulletin}\ }\textbf {\bibinfo {volume} {25}},\
  \bibinfo {pages} {52} (\bibinfo {year} {2000})}\BibitemShut {NoStop}%
\bibitem [{\citenamefont {Edwards}\ \emph {et~al.}(2004)\citenamefont
  {Edwards}, \citenamefont {Porch}, \citenamefont {Jones}, \citenamefont
  {Morgan},\ and\ \citenamefont {Perks}}]{Edwards2004}%
  \BibitemOpen
  \bibfield  {author} {\bibinfo {author} {\bibfnamefont {P.~P.}\ \bibnamefont
  {Edwards}}, \bibinfo {author} {\bibfnamefont {A.}~\bibnamefont {Porch}},
  \bibinfo {author} {\bibfnamefont {M.~O.}\ \bibnamefont {Jones}}, \bibinfo
  {author} {\bibfnamefont {D.~V.}\ \bibnamefont {Morgan}}, \ and\ \bibinfo
  {author} {\bibfnamefont {R.~M.}\ \bibnamefont {Perks}},\ }\href {\doibase
  10.1039/b408864f} {\bibfield  {journal} {\bibinfo  {journal} {Dalton
  Transactions}\ ,\ \bibinfo {pages} {2995}} (\bibinfo {year}
  {2004})}\BibitemShut {NoStop}%
\bibitem [{\citenamefont {Kumar}\ and\ \citenamefont {Zhou}(2010)}]{Kumar2010}%
  \BibitemOpen
  \bibfield  {author} {\bibinfo {author} {\bibfnamefont {A.}~\bibnamefont
  {Kumar}}\ and\ \bibinfo {author} {\bibfnamefont {C.}~\bibnamefont {Zhou}},\
  }\href {\doibase 10.1021/nn901903b} {\bibfield  {journal} {\bibinfo
  {journal} {ACS Nano}\ }\textbf {\bibinfo {volume} {4}},\ \bibinfo {pages}
  {11} (\bibinfo {year} {2010})}\BibitemShut {NoStop}%
\bibitem [{\citenamefont {Gao}\ \emph {et~al.}(2016)\citenamefont {Gao},
  \citenamefont {Kempa}, \citenamefont {Giersig}, \citenamefont {Akinoglu},
  \citenamefont {Han},\ and\ \citenamefont {Li}}]{Gao2016}%
  \BibitemOpen
  \bibfield  {author} {\bibinfo {author} {\bibfnamefont {J.}~\bibnamefont
  {Gao}}, \bibinfo {author} {\bibfnamefont {K.}~\bibnamefont {Kempa}}, \bibinfo
  {author} {\bibfnamefont {M.}~\bibnamefont {Giersig}}, \bibinfo {author}
  {\bibfnamefont {E.~M.}\ \bibnamefont {Akinoglu}}, \bibinfo {author}
  {\bibfnamefont {B.}~\bibnamefont {Han}}, \ and\ \bibinfo {author}
  {\bibfnamefont {R.}~\bibnamefont {Li}},\ }\href {\doibase
  10.1080/00018732.2016.1226804} {\bibfield  {journal} {\bibinfo  {journal}
  {Advances in Physics}\ }\textbf {\bibinfo {volume} {65}},\ \bibinfo {pages}
  {553} (\bibinfo {year} {2016})}\BibitemShut {NoStop}%
\bibitem [{\citenamefont {Layani}\ \emph {et~al.}(2014)\citenamefont {Layani},
  \citenamefont {Kamyshny},\ and\ \citenamefont {Magdassi}}]{Layani2014}%
  \BibitemOpen
  \bibfield  {author} {\bibinfo {author} {\bibfnamefont {M.}~\bibnamefont
  {Layani}}, \bibinfo {author} {\bibfnamefont {A.}~\bibnamefont {Kamyshny}}, \
  and\ \bibinfo {author} {\bibfnamefont {S.}~\bibnamefont {Magdassi}},\ }\href
  {\doibase 10.1039/C4NR00102H} {\bibfield  {journal} {\bibinfo  {journal}
  {Nanoscale}\ }\textbf {\bibinfo {volume} {6}},\ \bibinfo {pages} {5581}
  (\bibinfo {year} {2014})}\BibitemShut {NoStop}%
\bibitem [{\citenamefont {Lee}\ \emph {et~al.}(2017)\citenamefont {Lee},
  \citenamefont {Kim}, \citenamefont {Kang}, \citenamefont {Jang},
  \citenamefont {Kim}, \citenamefont {Lee},\ and\ \citenamefont
  {Kim}}]{Lee2017}%
  \BibitemOpen
  \bibfield  {author} {\bibinfo {author} {\bibfnamefont {W.-J.}\ \bibnamefont
  {Lee}}, \bibinfo {author} {\bibfnamefont {H.~J.}\ \bibnamefont {Kim}},
  \bibinfo {author} {\bibfnamefont {J.}~\bibnamefont {Kang}}, \bibinfo {author}
  {\bibfnamefont {D.~H.}\ \bibnamefont {Jang}}, \bibinfo {author}
  {\bibfnamefont {T.~H.}\ \bibnamefont {Kim}}, \bibinfo {author} {\bibfnamefont
  {J.~H.}\ \bibnamefont {Lee}}, \ and\ \bibinfo {author} {\bibfnamefont
  {K.~H.}\ \bibnamefont {Kim}},\ }\href {\doibase
  10.1146/annurev-matsci-070616-124109} {\bibfield  {journal} {\bibinfo
  {journal} {Annual Review of Materials Research}\ }\textbf {\bibinfo {volume}
  {47}},\ \bibinfo {pages} {391} (\bibinfo {year} {2017})}\BibitemShut
  {NoStop}%
\bibitem [{\citenamefont {Zhang}\ \emph {et~al.}(2015)\citenamefont {Zhang},
  \citenamefont {Zhang}, \citenamefont {Perkins},\ and\ \citenamefont
  {Zunger}}]{Zhang2015}%
  \BibitemOpen
  \bibfield  {author} {\bibinfo {author} {\bibfnamefont {X.}~\bibnamefont
  {Zhang}}, \bibinfo {author} {\bibfnamefont {L.}~\bibnamefont {Zhang}},
  \bibinfo {author} {\bibfnamefont {J.~D.}\ \bibnamefont {Perkins}}, \ and\
  \bibinfo {author} {\bibfnamefont {A.}~\bibnamefont {Zunger}},\ }\href
  {\doibase 10.1103/PhysRevLett.115.176602} {\bibfield  {journal} {\bibinfo
  {journal} {Physical Review Letters}\ }\textbf {\bibinfo {volume} {115}},\
  \bibinfo {pages} {176602} (\bibinfo {year} {2015})}\BibitemShut {NoStop}%
\bibitem [{\citenamefont {Fox}(2001)}]{Fox2001}%
  \BibitemOpen
  \bibfield  {author} {\bibinfo {author} {\bibfnamefont {A.~M.}\ \bibnamefont
  {Fox}},\ }\href {https://books.google.com/books?id=-5bVBbAoaGoC} {\emph
  {\bibinfo {title} {{Optical Properties of Solids}}}},\ Oxford master series
  in condensed matter physics\ (\bibinfo  {publisher} {Oxford University
  Press},\ \bibinfo {year} {2001})\BibitemShut {NoStop}%
\bibitem [{\citenamefont {Zhang}\ \emph {et~al.}(2016)\citenamefont {Zhang},
  \citenamefont {Zhou}, \citenamefont {Guo}, \citenamefont {Zhao},
  \citenamefont {Barnes}, \citenamefont {Zhang}, \citenamefont {Eaton},
  \citenamefont {Zheng}, \citenamefont {Brahlek}, \citenamefont {Haneef},
  \citenamefont {Podraza}, \citenamefont {Chan}, \citenamefont {Gopalan},
  \citenamefont {Rabe},\ and\ \citenamefont {Engel-Herbert}}]{Zhang2016}%
  \BibitemOpen
  \bibfield  {author} {\bibinfo {author} {\bibfnamefont {L.}~\bibnamefont
  {Zhang}}, \bibinfo {author} {\bibfnamefont {Y.}~\bibnamefont {Zhou}},
  \bibinfo {author} {\bibfnamefont {L.}~\bibnamefont {Guo}}, \bibinfo {author}
  {\bibfnamefont {W.}~\bibnamefont {Zhao}}, \bibinfo {author} {\bibfnamefont
  {A.}~\bibnamefont {Barnes}}, \bibinfo {author} {\bibfnamefont {H.-T.}\
  \bibnamefont {Zhang}}, \bibinfo {author} {\bibfnamefont {C.}~\bibnamefont
  {Eaton}}, \bibinfo {author} {\bibfnamefont {Y.}~\bibnamefont {Zheng}},
  \bibinfo {author} {\bibfnamefont {M.}~\bibnamefont {Brahlek}}, \bibinfo
  {author} {\bibfnamefont {H.~F.}\ \bibnamefont {Haneef}}, \bibinfo {author}
  {\bibfnamefont {N.~J.}\ \bibnamefont {Podraza}}, \bibinfo {author}
  {\bibfnamefont {M.~H.~W.}\ \bibnamefont {Chan}}, \bibinfo {author}
  {\bibfnamefont {V.}~\bibnamefont {Gopalan}}, \bibinfo {author} {\bibfnamefont
  {K.~M.}\ \bibnamefont {Rabe}}, \ and\ \bibinfo {author} {\bibfnamefont
  {R.}~\bibnamefont {Engel-Herbert}},\ }\href {\doibase 10.1038/nmat4493}
  {\bibfield  {journal} {\bibinfo  {journal} {Nature Materials}\ }\textbf
  {\bibinfo {volume} {15}},\ \bibinfo {pages} {204} (\bibinfo {year}
  {2016})}\BibitemShut {NoStop}%
\bibitem [{\citenamefont {Inoue}\ \emph {et~al.}(1998)\citenamefont {Inoue},
  \citenamefont {Goto}, \citenamefont {Makino}, \citenamefont {Hussey},\ and\
  \citenamefont {Ishikawa}}]{Inoue1998}%
  \BibitemOpen
  \bibfield  {author} {\bibinfo {author} {\bibfnamefont {I.~H.}\ \bibnamefont
  {Inoue}}, \bibinfo {author} {\bibfnamefont {O.}~\bibnamefont {Goto}},
  \bibinfo {author} {\bibfnamefont {H.}~\bibnamefont {Makino}}, \bibinfo
  {author} {\bibfnamefont {N.~E.}\ \bibnamefont {Hussey}}, \ and\ \bibinfo
  {author} {\bibfnamefont {M.}~\bibnamefont {Ishikawa}},\ }\href {\doibase
  10.1103/PhysRevB.58.4372} {\bibfield  {journal} {\bibinfo  {journal}
  {Physical Review B}\ }\textbf {\bibinfo {volume} {58}},\ \bibinfo {pages}
  {4372} (\bibinfo {year} {1998})}\BibitemShut {NoStop}%
\bibitem [{\citenamefont {Yoshida}\ \emph {et~al.}(2010)\citenamefont
  {Yoshida}, \citenamefont {Hashimoto}, \citenamefont {Takizawa}, \citenamefont
  {Fujimori}, \citenamefont {Kubota}, \citenamefont {Ono},\ and\ \citenamefont
  {Eisaki}}]{Yoshida2010}%
  \BibitemOpen
  \bibfield  {author} {\bibinfo {author} {\bibfnamefont {T.}~\bibnamefont
  {Yoshida}}, \bibinfo {author} {\bibfnamefont {M.}~\bibnamefont {Hashimoto}},
  \bibinfo {author} {\bibfnamefont {T.}~\bibnamefont {Takizawa}}, \bibinfo
  {author} {\bibfnamefont {A.}~\bibnamefont {Fujimori}}, \bibinfo {author}
  {\bibfnamefont {M.}~\bibnamefont {Kubota}}, \bibinfo {author} {\bibfnamefont
  {K.}~\bibnamefont {Ono}}, \ and\ \bibinfo {author} {\bibfnamefont
  {H.}~\bibnamefont {Eisaki}},\ }\href {\doibase 10.1103/PhysRevB.82.085119}
  {\bibfield  {journal} {\bibinfo  {journal} {Physical Review B}\ }\textbf
  {\bibinfo {volume} {82}},\ \bibinfo {pages} {85119} (\bibinfo {year}
  {2010})}\BibitemShut {NoStop}%
\bibitem [{\citenamefont {Bozovic}\ \emph {et~al.}(2002)\citenamefont
  {Bozovic}, \citenamefont {Logvenov}, \citenamefont {Belca}, \citenamefont
  {Narimbetov},\ and\ \citenamefont {Sveklo}}]{Bozovic2002}%
  \BibitemOpen
  \bibfield  {author} {\bibinfo {author} {\bibfnamefont {I.}~\bibnamefont
  {Bozovic}}, \bibinfo {author} {\bibfnamefont {G.}~\bibnamefont {Logvenov}},
  \bibinfo {author} {\bibfnamefont {I.}~\bibnamefont {Belca}}, \bibinfo
  {author} {\bibfnamefont {B.}~\bibnamefont {Narimbetov}}, \ and\ \bibinfo
  {author} {\bibfnamefont {I.}~\bibnamefont {Sveklo}},\ }\href {\doibase
  10.1103/PhysRevLett.89.107001} {\bibfield  {journal} {\bibinfo  {journal}
  {Physical Review Letters}\ }\textbf {\bibinfo {volume} {89}},\ \bibinfo
  {pages} {107001} (\bibinfo {year} {2002})}\BibitemShut {NoStop}%
\bibitem [{\citenamefont {Haeni}\ \emph {et~al.}(2004)\citenamefont {Haeni},
  \citenamefont {Irvin}, \citenamefont {Chang}, \citenamefont {Uecker},
  \citenamefont {Reiche}, \citenamefont {Li}, \citenamefont {Choudhury},
  \citenamefont {Tian}, \citenamefont {Hawley}, \citenamefont {Craigo},
  \citenamefont {Tagantsev}, \citenamefont {Pan}, \citenamefont {Streiffer},
  \citenamefont {Chen}, \citenamefont {Kirchoefer}, \citenamefont {Levy},\ and\
  \citenamefont {Schlom}}]{Haeni2004}%
  \BibitemOpen
  \bibfield  {author} {\bibinfo {author} {\bibfnamefont {J.~H.}\ \bibnamefont
  {Haeni}}, \bibinfo {author} {\bibfnamefont {P.}~\bibnamefont {Irvin}},
  \bibinfo {author} {\bibfnamefont {W.}~\bibnamefont {Chang}}, \bibinfo
  {author} {\bibfnamefont {R.}~\bibnamefont {Uecker}}, \bibinfo {author}
  {\bibfnamefont {P.}~\bibnamefont {Reiche}}, \bibinfo {author} {\bibfnamefont
  {Y.~L.}\ \bibnamefont {Li}}, \bibinfo {author} {\bibfnamefont
  {S.}~\bibnamefont {Choudhury}}, \bibinfo {author} {\bibfnamefont
  {W.}~\bibnamefont {Tian}}, \bibinfo {author} {\bibfnamefont {M.~E.}\
  \bibnamefont {Hawley}}, \bibinfo {author} {\bibfnamefont {B.}~\bibnamefont
  {Craigo}}, \bibinfo {author} {\bibfnamefont {A.~K.}\ \bibnamefont
  {Tagantsev}}, \bibinfo {author} {\bibfnamefont {X.~Q.}\ \bibnamefont {Pan}},
  \bibinfo {author} {\bibfnamefont {S.~K.}\ \bibnamefont {Streiffer}}, \bibinfo
  {author} {\bibfnamefont {L.~Q.}\ \bibnamefont {Chen}}, \bibinfo {author}
  {\bibfnamefont {S.~W.}\ \bibnamefont {Kirchoefer}}, \bibinfo {author}
  {\bibfnamefont {J.}~\bibnamefont {Levy}}, \ and\ \bibinfo {author}
  {\bibfnamefont {D.~G.}\ \bibnamefont {Schlom}},\ }\href {\doibase
  10.1038/nature02773} {\bibfield  {journal} {\bibinfo  {journal} {Nature}\
  }\textbf {\bibinfo {volume} {430}},\ \bibinfo {pages} {758} (\bibinfo {year}
  {2004})}\BibitemShut {NoStop}%
\bibitem [{\citenamefont {Choi}(2004)}]{Choi2004}%
  \BibitemOpen
  \bibfield  {author} {\bibinfo {author} {\bibfnamefont {K.~J.}\ \bibnamefont
  {Choi}},\ }\href {\doibase 10.1126/science.1103218} {\bibfield  {journal}
  {\bibinfo  {journal} {Science}\ }\textbf {\bibinfo {volume} {306}},\ \bibinfo
  {pages} {1005} (\bibinfo {year} {2004})}\BibitemShut {NoStop}%
\bibitem [{\citenamefont {Lee}\ and\ \citenamefont
  {Rabe}(2010)}]{Lee2010_rabe}%
  \BibitemOpen
  \bibfield  {author} {\bibinfo {author} {\bibfnamefont {J.~H.}\ \bibnamefont
  {Lee}}\ and\ \bibinfo {author} {\bibfnamefont {K.~M.}\ \bibnamefont {Rabe}},\
  }\href {\doibase 10.1103/PhysRevLett.104.207204} {\bibfield  {journal}
  {\bibinfo  {journal} {Physical Review Letters}\ }\textbf {\bibinfo {volume}
  {104}},\ \bibinfo {pages} {207204} (\bibinfo {year} {2010})}\BibitemShut
  {NoStop}%
\bibitem [{\citenamefont {Sclauzero}\ \emph {et~al.}(2016)\citenamefont
  {Sclauzero}, \citenamefont {Dymkowski},\ and\ \citenamefont
  {Ederer}}]{Sclauzero2016}%
  \BibitemOpen
  \bibfield  {author} {\bibinfo {author} {\bibfnamefont {G.}~\bibnamefont
  {Sclauzero}}, \bibinfo {author} {\bibfnamefont {K.}~\bibnamefont
  {Dymkowski}}, \ and\ \bibinfo {author} {\bibfnamefont {C.}~\bibnamefont
  {Ederer}},\ }\href {\doibase 10.1103/PhysRevB.94.245109} {\bibfield
  {journal} {\bibinfo  {journal} {Physical Review B}\ }\textbf {\bibinfo
  {volume} {94}},\ \bibinfo {pages} {245109} (\bibinfo {year}
  {2016})}\BibitemShut {NoStop}%
\bibitem [{\citenamefont {Beck}\ \emph {et~al.}(2018)\citenamefont {Beck},
  \citenamefont {Sclauzero}, \citenamefont {Chopra},\ and\ \citenamefont
  {Ederer}}]{Beck2018}%
  \BibitemOpen
  \bibfield  {author} {\bibinfo {author} {\bibfnamefont {S.}~\bibnamefont
  {Beck}}, \bibinfo {author} {\bibfnamefont {G.}~\bibnamefont {Sclauzero}},
  \bibinfo {author} {\bibfnamefont {U.}~\bibnamefont {Chopra}}, \ and\ \bibinfo
  {author} {\bibfnamefont {C.}~\bibnamefont {Ederer}},\ }\href {\doibase
  10.1103/PhysRevB.97.075107} {\bibfield  {journal} {\bibinfo  {journal}
  {Physical Review B}\ }\textbf {\bibinfo {volume} {97}},\ \bibinfo {pages}
  {75107} (\bibinfo {year} {2018})}\BibitemShut {NoStop}%
\bibitem [{\citenamefont {Dymkowski}\ and\ \citenamefont
  {Ederer}(2014)}]{Dymkowski2014}%
  \BibitemOpen
  \bibfield  {author} {\bibinfo {author} {\bibfnamefont {K.}~\bibnamefont
  {Dymkowski}}\ and\ \bibinfo {author} {\bibfnamefont {C.}~\bibnamefont
  {Ederer}},\ }\href {\doibase 10.1103/PhysRevB.89.161109} {\bibfield
  {journal} {\bibinfo  {journal} {Physical Review B}\ }\textbf {\bibinfo
  {volume} {89}},\ \bibinfo {pages} {161109(R)} (\bibinfo {year}
  {2014})}\BibitemShut {NoStop}%
\bibitem [{\citenamefont {Wadati}\ \emph {et~al.}(2014)\citenamefont {Wadati},
  \citenamefont {Mravlje}, \citenamefont {Yoshimatsu}, \citenamefont
  {Kumigashira}, \citenamefont {Oshima}, \citenamefont {Sugiyama},
  \citenamefont {Ikenaga}, \citenamefont {Fujimori}, \citenamefont {Georges},
  \citenamefont {Radetinac}, \citenamefont {Takahashi}, \citenamefont
  {Kawasaki},\ and\ \citenamefont {Tokura}}]{Wadati2014}%
  \BibitemOpen
  \bibfield  {author} {\bibinfo {author} {\bibfnamefont {H.}~\bibnamefont
  {Wadati}}, \bibinfo {author} {\bibfnamefont {J.}~\bibnamefont {Mravlje}},
  \bibinfo {author} {\bibfnamefont {K.}~\bibnamefont {Yoshimatsu}}, \bibinfo
  {author} {\bibfnamefont {H.}~\bibnamefont {Kumigashira}}, \bibinfo {author}
  {\bibfnamefont {M.}~\bibnamefont {Oshima}}, \bibinfo {author} {\bibfnamefont
  {T.}~\bibnamefont {Sugiyama}}, \bibinfo {author} {\bibfnamefont
  {E.}~\bibnamefont {Ikenaga}}, \bibinfo {author} {\bibfnamefont
  {A.}~\bibnamefont {Fujimori}}, \bibinfo {author} {\bibfnamefont
  {A.}~\bibnamefont {Georges}}, \bibinfo {author} {\bibfnamefont
  {A.}~\bibnamefont {Radetinac}}, \bibinfo {author} {\bibfnamefont {K.~S.}\
  \bibnamefont {Takahashi}}, \bibinfo {author} {\bibfnamefont {M.}~\bibnamefont
  {Kawasaki}}, \ and\ \bibinfo {author} {\bibfnamefont {Y.}~\bibnamefont
  {Tokura}},\ }\href {\doibase 10.1103/PhysRevB.90.205131} {\bibfield
  {journal} {\bibinfo  {journal} {Physical Review B}\ }\textbf {\bibinfo
  {volume} {90}},\ \bibinfo {pages} {205131} (\bibinfo {year}
  {2014})}\BibitemShut {NoStop}%
\bibitem [{\citenamefont {Kresse}\ and\ \citenamefont
  {Furthmuller}(1996)}]{VASP1}%
  \BibitemOpen
  \bibfield  {author} {\bibinfo {author} {\bibfnamefont {G.}~\bibnamefont
  {Kresse}}\ and\ \bibinfo {author} {\bibfnamefont {J.}~\bibnamefont
  {Furthmuller}},\ }\href@noop {} {\bibfield  {journal} {\bibinfo  {journal}
  {Computational Materials Science}\ }\textbf {\bibinfo {volume} {6}},\
  \bibinfo {pages} {15} (\bibinfo {year} {1996})}\BibitemShut {NoStop}%
\bibitem [{\citenamefont {Kresse}\ and\ \citenamefont
  {Furthm{\"{u}}ller}(1996)}]{VASP2}%
  \BibitemOpen
  \bibfield  {author} {\bibinfo {author} {\bibfnamefont {G.}~\bibnamefont
  {Kresse}}\ and\ \bibinfo {author} {\bibfnamefont {J.}~\bibnamefont
  {Furthm{\"{u}}ller}},\ }\href {\doibase 10.1103/PhysRevB.54.11169} {\bibfield
   {journal} {\bibinfo  {journal} {Physical Review B}\ }\textbf {\bibinfo
  {volume} {54}},\ \bibinfo {pages} {11169} (\bibinfo {year}
  {1996})}\BibitemShut {NoStop}%
\bibitem [{\citenamefont {Perdew}\ \emph {et~al.}(2008)\citenamefont {Perdew},
  \citenamefont {Ruzsinszky}, \citenamefont {Csonka}, \citenamefont {Vydrov},
  \citenamefont {Scuseria}, \citenamefont {Constantin}, \citenamefont {Zhou},\
  and\ \citenamefont {Burke}}]{PBEsol}%
  \BibitemOpen
  \bibfield  {author} {\bibinfo {author} {\bibfnamefont {J.~P.}\ \bibnamefont
  {Perdew}}, \bibinfo {author} {\bibfnamefont {A.}~\bibnamefont {Ruzsinszky}},
  \bibinfo {author} {\bibfnamefont {G.~I.}\ \bibnamefont {Csonka}}, \bibinfo
  {author} {\bibfnamefont {O.~A.}\ \bibnamefont {Vydrov}}, \bibinfo {author}
  {\bibfnamefont {G.~E.}\ \bibnamefont {Scuseria}}, \bibinfo {author}
  {\bibfnamefont {L.~A.}\ \bibnamefont {Constantin}}, \bibinfo {author}
  {\bibfnamefont {X.}~\bibnamefont {Zhou}}, \ and\ \bibinfo {author}
  {\bibfnamefont {K.}~\bibnamefont {Burke}},\ }\href {\doibase
  10.1103/PhysRevLett.100.136406} {\bibfield  {journal} {\bibinfo  {journal}
  {Physical Review Letters}\ }\textbf {\bibinfo {volume} {100}},\ \bibinfo
  {pages} {136406} (\bibinfo {year} {2008})}\BibitemShut {NoStop}%
\bibitem [{\citenamefont {Dudarev}\ \emph {et~al.}(1998)\citenamefont
  {Dudarev}, \citenamefont {Botton}, \citenamefont {Savrasov}, \citenamefont
  {Humphreys},\ and\ \citenamefont {Sutton}}]{DFTU:Dudarev}%
  \BibitemOpen
  \bibfield  {author} {\bibinfo {author} {\bibfnamefont {S.~L.}\ \bibnamefont
  {Dudarev}}, \bibinfo {author} {\bibfnamefont {G.~A.}\ \bibnamefont {Botton}},
  \bibinfo {author} {\bibfnamefont {S.~Y.}\ \bibnamefont {Savrasov}}, \bibinfo
  {author} {\bibfnamefont {C.~J.}\ \bibnamefont {Humphreys}}, \ and\ \bibinfo
  {author} {\bibfnamefont {A.~P.}\ \bibnamefont {Sutton}},\ }\href {\doibase
  10.1103/PhysRevB.57.1505} {\bibfield  {journal} {\bibinfo  {journal}
  {Physical Review B}\ }\textbf {\bibinfo {volume} {57}},\ \bibinfo {pages}
  {1505} (\bibinfo {year} {1998})}\BibitemShut {NoStop}%
\bibitem [{\citenamefont {Monkhorst}\ and\ \citenamefont
  {Pack}(1976)}]{Monkhorst1976}%
  \BibitemOpen
  \bibfield  {author} {\bibinfo {author} {\bibfnamefont {H.~J.}\ \bibnamefont
  {Monkhorst}}\ and\ \bibinfo {author} {\bibfnamefont {J.~D.}\ \bibnamefont
  {Pack}},\ }\href {\doibase 10.1103/PhysRevB.13.5188} {\bibfield  {journal}
  {\bibinfo  {journal} {Phys. Rev. B}\ }\textbf {\bibinfo {volume} {13}},\
  \bibinfo {pages} {5188} (\bibinfo {year} {1976})}\BibitemShut {NoStop}%
\bibitem [{\citenamefont {Blaha}\ \emph {et~al.}(2001)\citenamefont {Blaha},
  \citenamefont {K.}, \citenamefont {G.}, \citenamefont {D.},\ and\
  \citenamefont {J.}}]{WIEN2k}%
  \BibitemOpen
  \bibfield  {author} {\bibinfo {author} {\bibfnamefont {P.}~\bibnamefont
  {Blaha}}, \bibinfo {author} {\bibfnamefont {S.}~\bibnamefont {K.}}, \bibinfo
  {author} {\bibfnamefont {M.}~\bibnamefont {G.}}, \bibinfo {author}
  {\bibfnamefont {K.}~\bibnamefont {D.}}, \ and\ \bibinfo {author}
  {\bibfnamefont {L.}~\bibnamefont {J.}},\ }\href@noop {} {\emph {\bibinfo
  {title} {{WIEN2k: An Augmented Plane Wave Program for Calculating Crystal
  Properties}}}},\ \bibinfo {type} {Tech. Rep.}\ (\bibinfo {year}
  {2001})\BibitemShut {NoStop}%
\bibitem [{\citenamefont {Ambrosch-Draxl}\ and\ \citenamefont
  {Sofo}(2006)}]{Ambroschdraxl2006}%
  \BibitemOpen
  \bibfield  {author} {\bibinfo {author} {\bibfnamefont {C.}~\bibnamefont
  {Ambrosch-Draxl}}\ and\ \bibinfo {author} {\bibfnamefont {J.~O.}\
  \bibnamefont {Sofo}},\ }\href {\doibase 10.1016/j.cpc.2006.03.005} {\bibfield
   {journal} {\bibinfo  {journal} {Computer Physics Communications}\ }\textbf
  {\bibinfo {volume} {175}},\ \bibinfo {pages} {1} (\bibinfo {year}
  {2006})}\BibitemShut {NoStop}%
\bibitem [{Note1()}]{Note1}%
  \BibitemOpen
  \bibinfo {note} {In this context, the 'core electrons' include electrons in
  all the bands that do not cross the Fermi level, which would include core,
  semicore, and part of the valence electrons in the terminology of first
  principles methods such as linearized augmented plane-wave
  methods.}\BibitemShut {Stop}%
\bibitem [{\citenamefont {Grosso}\ and\ \citenamefont
  {Parravicini}(2000)}]{Grosso2000}%
  \BibitemOpen
  \bibfield  {author} {\bibinfo {author} {\bibfnamefont {G.}~\bibnamefont
  {Grosso}}\ and\ \bibinfo {author} {\bibfnamefont {G.~P.}\ \bibnamefont
  {Parravicini}},\ }\href {https://books.google.com/books?id=L5RrQbbvWn8C}
  {\emph {\bibinfo {title} {{Solid State Physics}}}}\ (\bibinfo  {publisher}
  {Elsevier Science},\ \bibinfo {year} {2000})\BibitemShut {NoStop}%
\bibitem [{\citenamefont {Haule}\ \emph {et~al.}(2010)\citenamefont {Haule},
  \citenamefont {Yee},\ and\ \citenamefont {Kim}}]{Haule2010}%
  \BibitemOpen
  \bibfield  {author} {\bibinfo {author} {\bibfnamefont {K.}~\bibnamefont
  {Haule}}, \bibinfo {author} {\bibfnamefont {C.-H.}\ \bibnamefont {Yee}}, \
  and\ \bibinfo {author} {\bibfnamefont {K.}~\bibnamefont {Kim}},\ }\href
  {\doibase 10.1103/PhysRevB.81.195107} {\bibfield  {journal} {\bibinfo
  {journal} {Physical Review B}\ }\textbf {\bibinfo {volume} {81}},\ \bibinfo
  {pages} {195107} (\bibinfo {year} {2010})}\BibitemShut {NoStop}%
\bibitem [{\citenamefont {Haule}(2018)}]{Haule2018}%
  \BibitemOpen
  \bibfield  {author} {\bibinfo {author} {\bibfnamefont {K.}~\bibnamefont
  {Haule}},\ }\href {\doibase 10.7566/JPSJ.87.041005} {\bibfield  {journal}
  {\bibinfo  {journal} {Journal of the Physical Society of Japan}\ }\textbf
  {\bibinfo {volume} {87}},\ \bibinfo {pages} {041005} (\bibinfo {year}
  {2018})}\BibitemShut {NoStop}%
\bibitem [{\citenamefont {Haule}(2007)}]{Haule2007}%
  \BibitemOpen
  \bibfield  {author} {\bibinfo {author} {\bibfnamefont {K.}~\bibnamefont
  {Haule}},\ }\href {\doibase 10.1103/PhysRevB.75.155113} {\bibfield  {journal}
  {\bibinfo  {journal} {Physical Review B}\ }\textbf {\bibinfo {volume} {75}},\
  \bibinfo {pages} {155113} (\bibinfo {year} {2007})}\BibitemShut {NoStop}%
\bibitem [{\citenamefont {Haule}(2015)}]{Haule2015_dc}%
  \BibitemOpen
  \bibfield  {author} {\bibinfo {author} {\bibfnamefont {K.}~\bibnamefont
  {Haule}},\ }\href {\doibase 10.1103/PhysRevLett.115.196403} {\bibfield
  {journal} {\bibinfo  {journal} {Physical Review Letters}\ }\textbf {\bibinfo
  {volume} {115}},\ \bibinfo {pages} {196403} (\bibinfo {year}
  {2015})}\BibitemShut {NoStop}%
\bibitem [{\citenamefont {Haule}\ \emph {et~al.}(2014)\citenamefont {Haule},
  \citenamefont {Birol},\ and\ \citenamefont {Kotliar}}]{Haule2014}%
  \BibitemOpen
  \bibfield  {author} {\bibinfo {author} {\bibfnamefont {K.}~\bibnamefont
  {Haule}}, \bibinfo {author} {\bibfnamefont {T.}~\bibnamefont {Birol}}, \ and\
  \bibinfo {author} {\bibfnamefont {G.}~\bibnamefont {Kotliar}},\ }\href
  {\doibase 10.1103/PhysRevB.90.075136} {\bibfield  {journal} {\bibinfo
  {journal} {Physical Review B}\ }\textbf {\bibinfo {volume} {90}},\ \bibinfo
  {pages} {075136} (\bibinfo {year} {2014})}\BibitemShut {NoStop}%
\bibitem [{\citenamefont {Haule}\ and\ \citenamefont
  {Pascut}(2016)}]{Haule2016Forces}%
  \BibitemOpen
  \bibfield  {author} {\bibinfo {author} {\bibfnamefont {K.}~\bibnamefont
  {Haule}}\ and\ \bibinfo {author} {\bibfnamefont {G.~L.}\ \bibnamefont
  {Pascut}},\ }\href {\doibase 10.1103/PhysRevB.94.195146} {\bibfield
  {journal} {\bibinfo  {journal} {Physical Review B}\ }\textbf {\bibinfo
  {volume} {94}},\ \bibinfo {pages} {195146} (\bibinfo {year}
  {2016})}\BibitemShut {NoStop}%
\bibitem [{\citenamefont {Haule}\ and\ \citenamefont
  {Birol}(2015)}]{Haule2015_energies}%
  \BibitemOpen
  \bibfield  {author} {\bibinfo {author} {\bibfnamefont {K.}~\bibnamefont
  {Haule}}\ and\ \bibinfo {author} {\bibfnamefont {T.}~\bibnamefont {Birol}},\
  }\href {\doibase 10.1103/PhysRevLett.115.256402} {\bibfield  {journal}
  {\bibinfo  {journal} {Physical Review Letters}\ }\textbf {\bibinfo {volume}
  {115}},\ \bibinfo {pages} {256402} (\bibinfo {year} {2015})}\BibitemShut
  {NoStop}%
\bibitem [{\citenamefont {Paul}\ and\ \citenamefont {Birol}(2019)}]{Paul2019}%
  \BibitemOpen
  \bibfield  {author} {\bibinfo {author} {\bibfnamefont {A.}~\bibnamefont
  {Paul}}\ and\ \bibinfo {author} {\bibfnamefont {T.}~\bibnamefont {Birol}},\
  }\href {\doibase 10.1146/annurev-matsci-070218-121825} {\bibfield  {journal}
  {\bibinfo  {journal} {Annual Review of Materials Research}\ }\textbf
  {\bibinfo {volume} {49}},\ \bibinfo {pages} {31} (\bibinfo {year}
  {2019})}\BibitemShut {NoStop}%
\bibitem [{\citenamefont {Han}\ \emph {et~al.}(2016)\citenamefont {Han},
  \citenamefont {Dang},\ and\ \citenamefont {Millis}}]{Han2016}%
  \BibitemOpen
  \bibfield  {author} {\bibinfo {author} {\bibfnamefont {Q.}~\bibnamefont
  {Han}}, \bibinfo {author} {\bibfnamefont {H.~T.}\ \bibnamefont {Dang}}, \
  and\ \bibinfo {author} {\bibfnamefont {A.~J.}\ \bibnamefont {Millis}},\
  }\href {\doibase 10.1103/PhysRevB.93.155103} {\bibfield  {journal} {\bibinfo
  {journal} {Physical Review B}\ }\textbf {\bibinfo {volume} {93}},\ \bibinfo
  {pages} {155103} (\bibinfo {year} {2016})}\BibitemShut {NoStop}%
\bibitem [{\citenamefont {Lufaso}\ and\ \citenamefont
  {Woodward}(2001)}]{Lufaso2001}%
  \BibitemOpen
  \bibfield  {author} {\bibinfo {author} {\bibfnamefont {M.~W.}\ \bibnamefont
  {Lufaso}}\ and\ \bibinfo {author} {\bibfnamefont {P.~M.}\ \bibnamefont
  {Woodward}},\ }\href {\doibase 10.1107/S0108768101015282} {\bibfield
  {journal} {\bibinfo  {journal} {Acta Crystallographica Section B: Structural
  Science}\ }\textbf {\bibinfo {volume} {57}},\ \bibinfo {pages} {725}
  (\bibinfo {year} {2001})}\BibitemShut {NoStop}%
\bibitem [{\citenamefont {Giannakopoulou}\ \emph {et~al.}(1995)\citenamefont
  {Giannakopoulou}, \citenamefont {Odier}, \citenamefont {Bassat},\ and\
  \citenamefont {Loup}}]{Giannakopoulou1995}%
  \BibitemOpen
  \bibfield  {author} {\bibinfo {author} {\bibfnamefont {V.}~\bibnamefont
  {Giannakopoulou}}, \bibinfo {author} {\bibfnamefont {P.}~\bibnamefont
  {Odier}}, \bibinfo {author} {\bibfnamefont {J.~M.}\ \bibnamefont {Bassat}}, \
  and\ \bibinfo {author} {\bibfnamefont {J.~P.}\ \bibnamefont {Loup}},\ }\href
  {\doibase 10.1016/0038-1098(94)00834-Y} {\bibfield  {journal} {\bibinfo
  {journal} {Solid State Communications}\ }\textbf {\bibinfo {volume} {93}},\
  \bibinfo {pages} {579} (\bibinfo {year} {1995})}\BibitemShut {NoStop}%
\bibitem [{\citenamefont {Pavarini}\ \emph {et~al.}(2004)\citenamefont
  {Pavarini}, \citenamefont {Biermann}, \citenamefont {Poteryaev},
  \citenamefont {Lichtenstein}, \citenamefont {Georges},\ and\ \citenamefont
  {Andersen}}]{Pavarini2004}%
  \BibitemOpen
  \bibfield  {author} {\bibinfo {author} {\bibfnamefont {E.}~\bibnamefont
  {Pavarini}}, \bibinfo {author} {\bibfnamefont {S.}~\bibnamefont {Biermann}},
  \bibinfo {author} {\bibfnamefont {A.}~\bibnamefont {Poteryaev}}, \bibinfo
  {author} {\bibfnamefont {A.~I.}\ \bibnamefont {Lichtenstein}}, \bibinfo
  {author} {\bibfnamefont {A.}~\bibnamefont {Georges}}, \ and\ \bibinfo
  {author} {\bibfnamefont {O.~K.}\ \bibnamefont {Andersen}},\ }\href {\doibase
  10.1103/PhysRevLett.92.176403} {\bibfield  {journal} {\bibinfo  {journal}
  {Physical Review Letters}\ }\textbf {\bibinfo {volume} {92}},\ \bibinfo
  {pages} {176403} (\bibinfo {year} {2004})}\BibitemShut {NoStop}%
\bibitem [{\citenamefont {Sakuma}\ \emph {et~al.}(2013)\citenamefont {Sakuma},
  \citenamefont {Werner},\ and\ \citenamefont {Aryasetiawan}}]{Sakuma2013}%
  \BibitemOpen
  \bibfield  {author} {\bibinfo {author} {\bibfnamefont {R.}~\bibnamefont
  {Sakuma}}, \bibinfo {author} {\bibfnamefont {P.}~\bibnamefont {Werner}}, \
  and\ \bibinfo {author} {\bibfnamefont {F.}~\bibnamefont {Aryasetiawan}},\
  }\href {\doibase 10.1103/PhysRevB.88.235110} {\bibfield  {journal} {\bibinfo
  {journal} {Physical Review B}\ }\textbf {\bibinfo {volume} {88}},\ \bibinfo
  {pages} {235110} (\bibinfo {year} {2013})}\BibitemShut {NoStop}%
\bibitem [{\citenamefont {Taranto}\ \emph {et~al.}(2013)\citenamefont
  {Taranto}, \citenamefont {Kaltak}, \citenamefont {Parragh}, \citenamefont
  {Sangiovanni}, \citenamefont {Kresse}, \citenamefont {Toschi},\ and\
  \citenamefont {Held}}]{Taranto2013}%
  \BibitemOpen
  \bibfield  {author} {\bibinfo {author} {\bibfnamefont {C.}~\bibnamefont
  {Taranto}}, \bibinfo {author} {\bibfnamefont {M.}~\bibnamefont {Kaltak}},
  \bibinfo {author} {\bibfnamefont {N.}~\bibnamefont {Parragh}}, \bibinfo
  {author} {\bibfnamefont {G.}~\bibnamefont {Sangiovanni}}, \bibinfo {author}
  {\bibfnamefont {G.}~\bibnamefont {Kresse}}, \bibinfo {author} {\bibfnamefont
  {A.}~\bibnamefont {Toschi}}, \ and\ \bibinfo {author} {\bibfnamefont
  {K.}~\bibnamefont {Held}},\ }\href {\doibase 10.1103/PhysRevB.88.165119}
  {\bibfield  {journal} {\bibinfo  {journal} {Physical Review B}\ }\textbf
  {\bibinfo {volume} {88}},\ \bibinfo {pages} {165119} (\bibinfo {year}
  {2013})}\BibitemShut {NoStop}%
\bibitem [{\citenamefont {Werner}\ and\ \citenamefont
  {Casula}(2016)}]{Werner2016}%
  \BibitemOpen
  \bibfield  {author} {\bibinfo {author} {\bibfnamefont {P.}~\bibnamefont
  {Werner}}\ and\ \bibinfo {author} {\bibfnamefont {M.}~\bibnamefont
  {Casula}},\ }\href {\doibase 10.1088/0953-8984/28/38/383001} {\bibfield
  {journal} {\bibinfo  {journal} {Journal of Physics: Condensed Matter}\
  }\textbf {\bibinfo {volume} {28}},\ \bibinfo {pages} {383001} (\bibinfo
  {year} {2016})}\BibitemShut {NoStop}%
\bibitem [{\citenamefont {Takizawa}\ \emph {et~al.}(2009)\citenamefont
  {Takizawa}, \citenamefont {Minohara}, \citenamefont {Kumigashira},
  \citenamefont {Toyota}, \citenamefont {Oshima}, \citenamefont {Wadati},
  \citenamefont {Yoshida}, \citenamefont {Fujimori}, \citenamefont {Lippmaa},
  \citenamefont {Kawasaki}, \citenamefont {Koinuma}, \citenamefont {Sordi},\
  and\ \citenamefont {Rozenberg}}]{Takizawa2009}%
  \BibitemOpen
  \bibfield  {author} {\bibinfo {author} {\bibfnamefont {M.}~\bibnamefont
  {Takizawa}}, \bibinfo {author} {\bibfnamefont {M.}~\bibnamefont {Minohara}},
  \bibinfo {author} {\bibfnamefont {H.}~\bibnamefont {Kumigashira}}, \bibinfo
  {author} {\bibfnamefont {D.}~\bibnamefont {Toyota}}, \bibinfo {author}
  {\bibfnamefont {M.}~\bibnamefont {Oshima}}, \bibinfo {author} {\bibfnamefont
  {H.}~\bibnamefont {Wadati}}, \bibinfo {author} {\bibfnamefont
  {T.}~\bibnamefont {Yoshida}}, \bibinfo {author} {\bibfnamefont
  {A.}~\bibnamefont {Fujimori}}, \bibinfo {author} {\bibfnamefont
  {M.}~\bibnamefont {Lippmaa}}, \bibinfo {author} {\bibfnamefont
  {M.}~\bibnamefont {Kawasaki}}, \bibinfo {author} {\bibfnamefont
  {H.}~\bibnamefont {Koinuma}}, \bibinfo {author} {\bibfnamefont
  {G.}~\bibnamefont {Sordi}}, \ and\ \bibinfo {author} {\bibfnamefont
  {M.}~\bibnamefont {Rozenberg}},\ }\href {\doibase 10.1103/PhysRevB.80.235104}
  {\bibfield  {journal} {\bibinfo  {journal} {Physical Review B}\ }\textbf
  {\bibinfo {volume} {80}},\ \bibinfo {pages} {235104} (\bibinfo {year}
  {2009})}\BibitemShut {NoStop}%
\bibitem [{\citenamefont {Yoshida}\ \emph {et~al.}(2016)\citenamefont
  {Yoshida}, \citenamefont {Kobayashi}, \citenamefont {Yoshimatsu},
  \citenamefont {Kumigashira},\ and\ \citenamefont {Fujimori}}]{Yoshida2016}%
  \BibitemOpen
  \bibfield  {author} {\bibinfo {author} {\bibfnamefont {T.}~\bibnamefont
  {Yoshida}}, \bibinfo {author} {\bibfnamefont {M.}~\bibnamefont {Kobayashi}},
  \bibinfo {author} {\bibfnamefont {K.}~\bibnamefont {Yoshimatsu}}, \bibinfo
  {author} {\bibfnamefont {H.}~\bibnamefont {Kumigashira}}, \ and\ \bibinfo
  {author} {\bibfnamefont {A.}~\bibnamefont {Fujimori}},\ }\href {\doibase
  10.1016/j.elspec.2015.11.012} {\bibfield  {journal} {\bibinfo  {journal}
  {Journal of Electron Spectroscopy and Related Phenomena}\ }\textbf {\bibinfo
  {volume} {208}},\ \bibinfo {pages} {11} (\bibinfo {year} {2016})}\BibitemShut
  {NoStop}%
\bibitem [{\citenamefont {Burganov}\ \emph {et~al.}(2016)\citenamefont
  {Burganov}, \citenamefont {Adamo}, \citenamefont {Mulder}, \citenamefont
  {Uchida}, \citenamefont {King}, \citenamefont {Harter}, \citenamefont {Shai},
  \citenamefont {Gibbs}, \citenamefont {Mackenzie}, \citenamefont {Uecker},
  \citenamefont {Bruetzam}, \citenamefont {Beasley}, \citenamefont {Fennie},
  \citenamefont {Schlom},\ and\ \citenamefont {Shen}}]{Burganov2016}%
  \BibitemOpen
  \bibfield  {author} {\bibinfo {author} {\bibfnamefont {B.}~\bibnamefont
  {Burganov}}, \bibinfo {author} {\bibfnamefont {C.}~\bibnamefont {Adamo}},
  \bibinfo {author} {\bibfnamefont {A.}~\bibnamefont {Mulder}}, \bibinfo
  {author} {\bibfnamefont {M.}~\bibnamefont {Uchida}}, \bibinfo {author}
  {\bibfnamefont {P.~D.~C.}\ \bibnamefont {King}}, \bibinfo {author}
  {\bibfnamefont {J.~W.}\ \bibnamefont {Harter}}, \bibinfo {author}
  {\bibfnamefont {D.~E.}\ \bibnamefont {Shai}}, \bibinfo {author}
  {\bibfnamefont {A.~S.}\ \bibnamefont {Gibbs}}, \bibinfo {author}
  {\bibfnamefont {A.~P.}\ \bibnamefont {Mackenzie}}, \bibinfo {author}
  {\bibfnamefont {R.}~\bibnamefont {Uecker}}, \bibinfo {author} {\bibfnamefont
  {M.}~\bibnamefont {Bruetzam}}, \bibinfo {author} {\bibfnamefont {M.~R.}\
  \bibnamefont {Beasley}}, \bibinfo {author} {\bibfnamefont {C.~J.}\
  \bibnamefont {Fennie}}, \bibinfo {author} {\bibfnamefont {D.~G.}\
  \bibnamefont {Schlom}}, \ and\ \bibinfo {author} {\bibfnamefont {K.~M.}\
  \bibnamefont {Shen}},\ }\href {\doibase 10.1103/PhysRevLett.116.197003}
  {\bibfield  {journal} {\bibinfo  {journal} {Physical Review Letters}\
  }\textbf {\bibinfo {volume} {116}},\ \bibinfo {pages} {197003} (\bibinfo
  {year} {2016})}\BibitemShut {NoStop}%
\bibitem [{\citenamefont {He}\ \emph {et~al.}(2012)\citenamefont {He},
  \citenamefont {Sanders}, \citenamefont {Gray}, \citenamefont {Wong},
  \citenamefont {Mehta},\ and\ \citenamefont {Suzuki}}]{He2012}%
  \BibitemOpen
  \bibfield  {author} {\bibinfo {author} {\bibfnamefont {C.}~\bibnamefont
  {He}}, \bibinfo {author} {\bibfnamefont {T.~D.}\ \bibnamefont {Sanders}},
  \bibinfo {author} {\bibfnamefont {M.~T.}\ \bibnamefont {Gray}}, \bibinfo
  {author} {\bibfnamefont {F.~J.}\ \bibnamefont {Wong}}, \bibinfo {author}
  {\bibfnamefont {V.~V.}\ \bibnamefont {Mehta}}, \ and\ \bibinfo {author}
  {\bibfnamefont {Y.}~\bibnamefont {Suzuki}},\ }\href {\doibase
  10.1103/PhysRevB.86.081401} {\bibfield  {journal} {\bibinfo  {journal}
  {Physical Review B - Condensed Matter and Materials Physics}\ }\textbf
  {\bibinfo {volume} {86}},\ \bibinfo {pages} {081401(R)} (\bibinfo {year}
  {2012})}\BibitemShut {NoStop}%
\bibitem [{\citenamefont {Woodward}(1997{\natexlab{a}})}]{Woodward1997a}%
  \BibitemOpen
  \bibfield  {author} {\bibinfo {author} {\bibfnamefont {P.~M.}\ \bibnamefont
  {Woodward}},\ }\href {\doibase 10.1107/S0108768196010713} {\bibfield
  {journal} {\bibinfo  {journal} {Acta Crystallographica Section B Structural
  Science}\ }\textbf {\bibinfo {volume} {53}},\ \bibinfo {pages} {32} (\bibinfo
  {year} {1997}{\natexlab{a}})}\BibitemShut {NoStop}%
\bibitem [{\citenamefont {Woodward}(1997{\natexlab{b}})}]{Woodward1997b}%
  \BibitemOpen
  \bibfield  {author} {\bibinfo {author} {\bibfnamefont {P.~M.}\ \bibnamefont
  {Woodward}},\ }\href {\doibase 10.1107/S0108768196012050} {\bibfield
  {journal} {\bibinfo  {journal} {Acta Crystallographica Section B Structural
  Science}\ }\textbf {\bibinfo {volume} {53}},\ \bibinfo {pages} {44} (\bibinfo
  {year} {1997}{\natexlab{b}})}\BibitemShut {NoStop}%
\bibitem [{\citenamefont {Glazer}(1972)}]{Glazer1972}%
  \BibitemOpen
  \bibfield  {author} {\bibinfo {author} {\bibfnamefont {A.~M.}\ \bibnamefont
  {Glazer}},\ }\href {\doibase 10.1107/S0567740872007976} {\bibfield  {journal}
  {\bibinfo  {journal} {Acta Crystallographica Section B Structural
  Crystallography and Crystal Chemistry}\ }\textbf {\bibinfo {volume} {28}},\
  \bibinfo {pages} {3384} (\bibinfo {year} {1972})}\BibitemShut {NoStop}%
\bibitem [{\citenamefont {Nekrasov}\ \emph {et~al.}(2005)\citenamefont
  {Nekrasov}, \citenamefont {Keller}, \citenamefont {Kondakov}, \citenamefont
  {Kozhevnikov}, \citenamefont {Pruschke}, \citenamefont {Held}, \citenamefont
  {Vollhardt},\ and\ \citenamefont {Anisimov}}]{Nekrasov2005}%
  \BibitemOpen
  \bibfield  {author} {\bibinfo {author} {\bibfnamefont {I.~A.}\ \bibnamefont
  {Nekrasov}}, \bibinfo {author} {\bibfnamefont {G.}~\bibnamefont {Keller}},
  \bibinfo {author} {\bibfnamefont {D.~E.}\ \bibnamefont {Kondakov}}, \bibinfo
  {author} {\bibfnamefont {A.~V.}\ \bibnamefont {Kozhevnikov}}, \bibinfo
  {author} {\bibfnamefont {T.}~\bibnamefont {Pruschke}}, \bibinfo {author}
  {\bibfnamefont {K.}~\bibnamefont {Held}}, \bibinfo {author} {\bibfnamefont
  {D.}~\bibnamefont {Vollhardt}}, \ and\ \bibinfo {author} {\bibfnamefont
  {V.~I.}\ \bibnamefont {Anisimov}},\ }\href {\doibase
  10.1103/PhysRevB.72.155106} {\bibfield  {journal} {\bibinfo  {journal}
  {Physical Review B}\ }\textbf {\bibinfo {volume} {72}},\ \bibinfo {pages}
  {155106} (\bibinfo {year} {2005})}\BibitemShut {NoStop}%
\bibitem [{\citenamefont {Yang}\ \emph {et~al.}(2012)\citenamefont {Yang},
  \citenamefont {Ren}, \citenamefont {Wang},\ and\ \citenamefont
  {Bellaiche}}]{Yang2012_ETO}%
  \BibitemOpen
  \bibfield  {author} {\bibinfo {author} {\bibfnamefont {Y.}~\bibnamefont
  {Yang}}, \bibinfo {author} {\bibfnamefont {W.}~\bibnamefont {Ren}}, \bibinfo
  {author} {\bibfnamefont {D.}~\bibnamefont {Wang}}, \ and\ \bibinfo {author}
  {\bibfnamefont {L.}~\bibnamefont {Bellaiche}},\ }\href {\doibase
  10.1103/PhysRevLett.109.267602} {\bibfield  {journal} {\bibinfo  {journal}
  {Physical Review Letters}\ }\textbf {\bibinfo {volume} {109}},\ \bibinfo
  {pages} {267602} (\bibinfo {year} {2012})}\BibitemShut {NoStop}%
\bibitem [{\citenamefont {Birol}\ and\ \citenamefont
  {Fennie}(2013)}]{Birol2013_ETO}%
  \BibitemOpen
  \bibfield  {author} {\bibinfo {author} {\bibfnamefont {T.}~\bibnamefont
  {Birol}}\ and\ \bibinfo {author} {\bibfnamefont {C.~J.}\ \bibnamefont
  {Fennie}},\ }\href {\doibase 10.1103/PhysRevB.88.094103} {\bibfield
  {journal} {\bibinfo  {journal} {Physical Review B}\ }\textbf {\bibinfo
  {volume} {88}},\ \bibinfo {pages} {094103} (\bibinfo {year}
  {2013})}\BibitemShut {NoStop}%
\bibitem [{\citenamefont {Lu}\ \emph {et~al.}(2015)\citenamefont {Lu},
  \citenamefont {Song}, \citenamefont {Yang}, \citenamefont {Ding},
  \citenamefont {Chow},\ and\ \citenamefont {Chen}}]{Lu2015}%
  \BibitemOpen
  \bibfield  {author} {\bibinfo {author} {\bibfnamefont {W.}~\bibnamefont
  {Lu}}, \bibinfo {author} {\bibfnamefont {W.}~\bibnamefont {Song}}, \bibinfo
  {author} {\bibfnamefont {P.}~\bibnamefont {Yang}}, \bibinfo {author}
  {\bibfnamefont {J.}~\bibnamefont {Ding}}, \bibinfo {author} {\bibfnamefont
  {G.~M.}\ \bibnamefont {Chow}}, \ and\ \bibinfo {author} {\bibfnamefont
  {J.}~\bibnamefont {Chen}},\ }\href@noop {} {\bibfield  {journal} {\bibinfo
  {journal} {Scientific reports}\ }\textbf {\bibinfo {volume} {5}},\ \bibinfo
  {pages} {10245} (\bibinfo {year} {2015})}\BibitemShut {NoStop}%
\bibitem [{\citenamefont {Wong}\ \emph {et~al.}(2010)\citenamefont {Wong},
  \citenamefont {Baek}, \citenamefont {Chopdekar}, \citenamefont {Mehta},
  \citenamefont {Jang}, \citenamefont {Eom},\ and\ \citenamefont
  {Suzuki}}]{Wong2010_LTO}%
  \BibitemOpen
  \bibfield  {author} {\bibinfo {author} {\bibfnamefont {F.~J.}\ \bibnamefont
  {Wong}}, \bibinfo {author} {\bibfnamefont {S.-H.}\ \bibnamefont {Baek}},
  \bibinfo {author} {\bibfnamefont {R.~V.}\ \bibnamefont {Chopdekar}}, \bibinfo
  {author} {\bibfnamefont {V.~V.}\ \bibnamefont {Mehta}}, \bibinfo {author}
  {\bibfnamefont {H.-W.}\ \bibnamefont {Jang}}, \bibinfo {author}
  {\bibfnamefont {C.-B.}\ \bibnamefont {Eom}}, \ and\ \bibinfo {author}
  {\bibfnamefont {Y.}~\bibnamefont {Suzuki}},\ }\href {\doibase
  10.1103/PhysRevB.81.161101} {\bibfield  {journal} {\bibinfo  {journal}
  {Physical Review B}\ }\textbf {\bibinfo {volume} {81}},\ \bibinfo {pages}
  {161101(R)} (\bibinfo {year} {2010})}\BibitemShut {NoStop}%
\bibitem [{Note2()}]{Note2}%
  \BibitemOpen
  \bibinfo {note} {Part of the reason for this transition is the better strain
  accommodation of the $c$ axis of the Pnma cell, which is of different length
  than the other two axes. Other examples of the same phenomenon include
  SrSnO$_3$ and CaTiO$_3$.\cite {Wang2018, Eklund2009}}\BibitemShut {NoStop}%
\bibitem [{\citenamefont {Yin}\ \emph {et~al.}(2011)\citenamefont {Yin},
  \citenamefont {Haule},\ and\ \citenamefont {Kotliar}}]{Yin2011}%
  \BibitemOpen
  \bibfield  {author} {\bibinfo {author} {\bibfnamefont {Z.~P.}\ \bibnamefont
  {Yin}}, \bibinfo {author} {\bibfnamefont {K.}~\bibnamefont {Haule}}, \ and\
  \bibinfo {author} {\bibfnamefont {G.}~\bibnamefont {Kotliar}},\ }\href
  {\doibase 10.1038/nmat3120} {\bibfield  {journal} {\bibinfo  {journal}
  {Nature Materials}\ }\textbf {\bibinfo {volume} {10}},\ \bibinfo {pages}
  {932} (\bibinfo {year} {2011})}\BibitemShut {NoStop}%
\bibitem [{\citenamefont {Georges}\ \emph {et~al.}(2013)\citenamefont
  {Georges}, \citenamefont {de' Medici},\ and\ \citenamefont
  {Mravlje}}]{Georges2013}%
  \BibitemOpen
  \bibfield  {author} {\bibinfo {author} {\bibfnamefont {A.}~\bibnamefont
  {Georges}}, \bibinfo {author} {\bibfnamefont {L.}~\bibnamefont {de' Medici}},
  \ and\ \bibinfo {author} {\bibfnamefont {J.}~\bibnamefont {Mravlje}},\ }\href
  {\doibase 10.1146/annurev-conmatphys-020911-125045} {\bibfield  {journal}
  {\bibinfo  {journal} {Annual Review of Condensed Matter Physics}\ }\textbf
  {\bibinfo {volume} {4}},\ \bibinfo {pages} {137} (\bibinfo {year}
  {2013})}\BibitemShut {NoStop}%
\bibitem [{\citenamefont {Hannerz}\ \emph {et~al.}(1999)\citenamefont
  {Hannerz}, \citenamefont {Svensson}, \citenamefont {Istomin},\ and\
  \citenamefont {D'yachenko}}]{Hannerz1999}%
  \BibitemOpen
  \bibfield  {author} {\bibinfo {author} {\bibfnamefont {H.}~\bibnamefont
  {Hannerz}}, \bibinfo {author} {\bibfnamefont {G.}~\bibnamefont {Svensson}},
  \bibinfo {author} {\bibfnamefont {S.}~\bibnamefont {Istomin}}, \ and\
  \bibinfo {author} {\bibfnamefont {O.}~\bibnamefont {D'yachenko}},\ }\href
  {\doibase 10.1006/jssc.1999.8357} {\bibfield  {journal} {\bibinfo  {journal}
  {Journal of Solid State Chemistry}\ }\textbf {\bibinfo {volume} {147}},\
  \bibinfo {pages} {421} (\bibinfo {year} {1999})}\BibitemShut {NoStop}%
\bibitem [{\citenamefont {Macquart}\ \emph
  {et~al.}(2010{\natexlab{a}})\citenamefont {Macquart}, \citenamefont
  {Kennedy},\ and\ \citenamefont {Avdeev}}]{Macquart2010_SrNbO3}%
  \BibitemOpen
  \bibfield  {author} {\bibinfo {author} {\bibfnamefont {R.~B.}\ \bibnamefont
  {Macquart}}, \bibinfo {author} {\bibfnamefont {B.~J.}\ \bibnamefont
  {Kennedy}}, \ and\ \bibinfo {author} {\bibfnamefont {M.}~\bibnamefont
  {Avdeev}},\ }\href {\doibase 10.1016/j.jssc.2010.08.001} {\bibfield
  {journal} {\bibinfo  {journal} {Journal of Solid State Chemistry}\ }\textbf
  {\bibinfo {volume} {183}},\ \bibinfo {pages} {2400} (\bibinfo {year}
  {2010}{\natexlab{a}})}\BibitemShut {NoStop}%
\bibitem [{\citenamefont {Peng}\ \emph {et~al.}(1998)\citenamefont {Peng},
  \citenamefont {Irvine},\ and\ \citenamefont {Fitzgerald}}]{Peng1998}%
  \BibitemOpen
  \bibfield  {author} {\bibinfo {author} {\bibfnamefont {N.}~\bibnamefont
  {Peng}}, \bibinfo {author} {\bibfnamefont {J.~T.~S.}\ \bibnamefont {Irvine}},
  \ and\ \bibinfo {author} {\bibfnamefont {A.~G.}\ \bibnamefont {Fitzgerald}},\
  }\href {\doibase 10.1039/a708290h} {\bibfield  {journal} {\bibinfo  {journal}
  {Journal of Materials Chemistry}\ }\textbf {\bibinfo {volume} {8}},\ \bibinfo
  {pages} {1033} (\bibinfo {year} {1998})}\BibitemShut {NoStop}%
\bibitem [{\citenamefont {Xu}\ \emph {et~al.}(2012)\citenamefont {Xu},
  \citenamefont {Randorn}, \citenamefont {Efstathiou},\ and\ \citenamefont
  {Irvine}}]{Xu2012}%
  \BibitemOpen
  \bibfield  {author} {\bibinfo {author} {\bibfnamefont {X.}~\bibnamefont
  {Xu}}, \bibinfo {author} {\bibfnamefont {C.}~\bibnamefont {Randorn}},
  \bibinfo {author} {\bibfnamefont {P.}~\bibnamefont {Efstathiou}}, \ and\
  \bibinfo {author} {\bibfnamefont {J.~T.~S.}\ \bibnamefont {Irvine}},\ }\href
  {\doibase 10.1038/nmat3312} {\bibfield  {journal} {\bibinfo  {journal}
  {Nature Materials}\ }\textbf {\bibinfo {volume} {11}},\ \bibinfo {pages}
  {595} (\bibinfo {year} {2012})}\BibitemShut {NoStop}%
\bibitem [{\citenamefont {Efstathiou}\ \emph {et~al.}(2013)\citenamefont
  {Efstathiou}, \citenamefont {Xu}, \citenamefont {M{\'{e}}nard},\ and\
  \citenamefont {Irvine}}]{Efstathiou2013}%
  \BibitemOpen
  \bibfield  {author} {\bibinfo {author} {\bibfnamefont {P.}~\bibnamefont
  {Efstathiou}}, \bibinfo {author} {\bibfnamefont {X.}~\bibnamefont {Xu}},
  \bibinfo {author} {\bibfnamefont {H.}~\bibnamefont {M{\'{e}}nard}}, \ and\
  \bibinfo {author} {\bibfnamefont {J.~T.~S.}\ \bibnamefont {Irvine}},\ }\href
  {\doibase 10.1039/c3dt32064b} {\bibfield  {journal} {\bibinfo  {journal}
  {Dalton Transactions}\ }\textbf {\bibinfo {volume} {42}},\ \bibinfo {pages}
  {7880} (\bibinfo {year} {2013})}\BibitemShut {NoStop}%
\bibitem [{\citenamefont {Wan}\ \emph {et~al.}(2017)\citenamefont {Wan},
  \citenamefont {Zhao}, \citenamefont {Cai}, \citenamefont {Asmara},
  \citenamefont {Huang}, \citenamefont {Chen}, \citenamefont {Hong},
  \citenamefont {Yin}, \citenamefont {Nelson}, \citenamefont {Motapothula},
  \citenamefont {Yan}, \citenamefont {Xiang}, \citenamefont {Chi},
  \citenamefont {Zheng}, \citenamefont {Chen}, \citenamefont {Xu},
  \citenamefont {Ariando}, \citenamefont {Rusydi}, \citenamefont {Minor},
  \citenamefont {Breese}, \citenamefont {Sherburne}, \citenamefont {Asta},
  \citenamefont {Xu},\ and\ \citenamefont {Venkatesan}}]{Wan2017}%
  \BibitemOpen
  \bibfield  {author} {\bibinfo {author} {\bibfnamefont {D.~Y.}\ \bibnamefont
  {Wan}}, \bibinfo {author} {\bibfnamefont {Y.~L.}\ \bibnamefont {Zhao}},
  \bibinfo {author} {\bibfnamefont {Y.}~\bibnamefont {Cai}}, \bibinfo {author}
  {\bibfnamefont {T.~C.}\ \bibnamefont {Asmara}}, \bibinfo {author}
  {\bibfnamefont {Z.}~\bibnamefont {Huang}}, \bibinfo {author} {\bibfnamefont
  {J.~Q.}\ \bibnamefont {Chen}}, \bibinfo {author} {\bibfnamefont
  {J.}~\bibnamefont {Hong}}, \bibinfo {author} {\bibfnamefont {S.~M.}\
  \bibnamefont {Yin}}, \bibinfo {author} {\bibfnamefont {C.~T.}\ \bibnamefont
  {Nelson}}, \bibinfo {author} {\bibfnamefont {M.~R.}\ \bibnamefont
  {Motapothula}}, \bibinfo {author} {\bibfnamefont {B.~X.}\ \bibnamefont
  {Yan}}, \bibinfo {author} {\bibfnamefont {D.}~\bibnamefont {Xiang}}, \bibinfo
  {author} {\bibfnamefont {X.}~\bibnamefont {Chi}}, \bibinfo {author}
  {\bibfnamefont {H.}~\bibnamefont {Zheng}}, \bibinfo {author} {\bibfnamefont
  {W.}~\bibnamefont {Chen}}, \bibinfo {author} {\bibfnamefont {R.}~\bibnamefont
  {Xu}}, \bibinfo {author} {\bibnamefont {Ariando}}, \bibinfo {author}
  {\bibfnamefont {A.}~\bibnamefont {Rusydi}}, \bibinfo {author} {\bibfnamefont
  {A.~M.}\ \bibnamefont {Minor}}, \bibinfo {author} {\bibfnamefont {M.~B.~H.}\
  \bibnamefont {Breese}}, \bibinfo {author} {\bibfnamefont {M.}~\bibnamefont
  {Sherburne}}, \bibinfo {author} {\bibfnamefont {M.}~\bibnamefont {Asta}},
  \bibinfo {author} {\bibfnamefont {Q.-H.}\ \bibnamefont {Xu}}, \ and\ \bibinfo
  {author} {\bibfnamefont {T.}~\bibnamefont {Venkatesan}},\ }\href {\doibase
  10.1038/ncomms15070} {\bibfield  {journal} {\bibinfo  {journal} {Nature
  Communications}\ }\textbf {\bibinfo {volume} {8}},\ \bibinfo {pages} {15070}
  (\bibinfo {year} {2017})}\BibitemShut {NoStop}%
\bibitem [{Note3()}]{Note3}%
  \BibitemOpen
  \bibinfo {note} {While there are lower energy Nb--d to Nb--d transitions,
  they give rise to only a small absorptivity peak around $\sim
  2.7$~eV.}\BibitemShut {Stop}%
\bibitem [{\citenamefont {Oka}\ \emph {et~al.}(2015)\citenamefont {Oka},
  \citenamefont {Hirose}, \citenamefont {Nakao}, \citenamefont {Fukumura},\
  and\ \citenamefont {Hasegawa}}]{Oka2015}%
  \BibitemOpen
  \bibfield  {author} {\bibinfo {author} {\bibfnamefont {D.}~\bibnamefont
  {Oka}}, \bibinfo {author} {\bibfnamefont {Y.}~\bibnamefont {Hirose}},
  \bibinfo {author} {\bibfnamefont {S.}~\bibnamefont {Nakao}}, \bibinfo
  {author} {\bibfnamefont {T.}~\bibnamefont {Fukumura}}, \ and\ \bibinfo
  {author} {\bibfnamefont {T.}~\bibnamefont {Hasegawa}},\ }\href {\doibase
  10.1103/PhysRevB.92.205102} {\bibfield  {journal} {\bibinfo  {journal}
  {Physical Review B}\ }\textbf {\bibinfo {volume} {92}},\ \bibinfo {pages}
  {205102} (\bibinfo {year} {2015})}\BibitemShut {NoStop}%
\bibitem [{\citenamefont {Puchkov}\ \emph {et~al.}(1998)\citenamefont
  {Puchkov}, \citenamefont {Schabel}, \citenamefont {Basov}, \citenamefont
  {Startseva}, \citenamefont {Cao}, \citenamefont {Timusk},\ and\ \citenamefont
  {Shen}}]{Shen1998}%
  \BibitemOpen
  \bibfield  {author} {\bibinfo {author} {\bibfnamefont {A.~V.}\ \bibnamefont
  {Puchkov}}, \bibinfo {author} {\bibfnamefont {M.~C.}\ \bibnamefont
  {Schabel}}, \bibinfo {author} {\bibfnamefont {D.~N.}\ \bibnamefont {Basov}},
  \bibinfo {author} {\bibfnamefont {T.}~\bibnamefont {Startseva}}, \bibinfo
  {author} {\bibfnamefont {G.}~\bibnamefont {Cao}}, \bibinfo {author}
  {\bibfnamefont {T.}~\bibnamefont {Timusk}}, \ and\ \bibinfo {author}
  {\bibfnamefont {Z.-X.}\ \bibnamefont {Shen}},\ }\href {\doibase
  10.1103/PhysRevLett.81.2747} {\bibfield  {journal} {\bibinfo  {journal}
  {Physical Review Letters}\ }\textbf {\bibinfo {volume} {81}},\ \bibinfo
  {pages} {2747} (\bibinfo {year} {1998})}\BibitemShut {NoStop}%
\bibitem [{\citenamefont {Kim}\ \emph {et~al.}(2012)\citenamefont {Kim},
  \citenamefont {Choi}, \citenamefont {Kim}, \citenamefont {Mitchell},
  \citenamefont {Jackeli}, \citenamefont {Daghofer}, \citenamefont {van~den
  Brink}, \citenamefont {Khaliullin},\ and\ \citenamefont {Kim}}]{Kim2012}%
  \BibitemOpen
  \bibfield  {author} {\bibinfo {author} {\bibfnamefont {J.~W.}\ \bibnamefont
  {Kim}}, \bibinfo {author} {\bibfnamefont {Y.}~\bibnamefont {Choi}}, \bibinfo
  {author} {\bibfnamefont {J.}~\bibnamefont {Kim}}, \bibinfo {author}
  {\bibfnamefont {J.~F.}\ \bibnamefont {Mitchell}}, \bibinfo {author}
  {\bibfnamefont {G.}~\bibnamefont {Jackeli}}, \bibinfo {author} {\bibfnamefont
  {M.}~\bibnamefont {Daghofer}}, \bibinfo {author} {\bibfnamefont
  {J.}~\bibnamefont {van~den Brink}}, \bibinfo {author} {\bibfnamefont
  {G.}~\bibnamefont {Khaliullin}}, \ and\ \bibinfo {author} {\bibfnamefont
  {B.~J.}\ \bibnamefont {Kim}},\ }\href {\doibase
  10.1103/PhysRevLett.109.037204} {\bibfield  {journal} {\bibinfo  {journal}
  {Physical Review Letters}\ }\textbf {\bibinfo {volume} {109}},\ \bibinfo
  {pages} {037204} (\bibinfo {year} {2012})}\BibitemShut {NoStop}%
\bibitem [{\citenamefont {Lee}\ \emph {et~al.}(2013)\citenamefont {Lee},
  \citenamefont {Orloff}, \citenamefont {Birol}, \citenamefont {Zhu},
  \citenamefont {Goian}, \citenamefont {Rocas}, \citenamefont {Haislmaier},
  \citenamefont {Vlahos}, \citenamefont {Mundy}, \citenamefont {Kourkoutis},
  \citenamefont {Nie}, \citenamefont {Biegalski}, \citenamefont {Zhang},
  \citenamefont {Bernhagen}, \citenamefont {Benedek}, \citenamefont {Kim},
  \citenamefont {Brock}, \citenamefont {Uecker}, \citenamefont {Xi},
  \citenamefont {Gopalan}, \citenamefont {Nuzhnyy}, \citenamefont {Kamba},
  \citenamefont {Muller}, \citenamefont {Takeuchi}, \citenamefont {Booth},
  \citenamefont {Fennie},\ and\ \citenamefont {Schlom}}]{Lee2013Birol}%
  \BibitemOpen
  \bibfield  {author} {\bibinfo {author} {\bibfnamefont {C.-H.~H.}\
  \bibnamefont {Lee}}, \bibinfo {author} {\bibfnamefont {N.~D.}\ \bibnamefont
  {Orloff}}, \bibinfo {author} {\bibfnamefont {T.}~\bibnamefont {Birol}},
  \bibinfo {author} {\bibfnamefont {Y.}~\bibnamefont {Zhu}}, \bibinfo {author}
  {\bibfnamefont {V.}~\bibnamefont {Goian}}, \bibinfo {author} {\bibfnamefont
  {E.}~\bibnamefont {Rocas}}, \bibinfo {author} {\bibfnamefont
  {R.}~\bibnamefont {Haislmaier}}, \bibinfo {author} {\bibfnamefont
  {E.}~\bibnamefont {Vlahos}}, \bibinfo {author} {\bibfnamefont {J.~A.}\
  \bibnamefont {Mundy}}, \bibinfo {author} {\bibfnamefont {L.~F.}\ \bibnamefont
  {Kourkoutis}}, \bibinfo {author} {\bibfnamefont {Y.}~\bibnamefont {Nie}},
  \bibinfo {author} {\bibfnamefont {M.~D.}\ \bibnamefont {Biegalski}}, \bibinfo
  {author} {\bibfnamefont {J.}~\bibnamefont {Zhang}}, \bibinfo {author}
  {\bibfnamefont {M.}~\bibnamefont {Bernhagen}}, \bibinfo {author}
  {\bibfnamefont {N.~A.}\ \bibnamefont {Benedek}}, \bibinfo {author}
  {\bibfnamefont {Y.}~\bibnamefont {Kim}}, \bibinfo {author} {\bibfnamefont
  {J.~D.}\ \bibnamefont {Brock}}, \bibinfo {author} {\bibfnamefont
  {R.}~\bibnamefont {Uecker}}, \bibinfo {author} {\bibfnamefont {X.~X.}\
  \bibnamefont {Xi}}, \bibinfo {author} {\bibfnamefont {V.}~\bibnamefont
  {Gopalan}}, \bibinfo {author} {\bibfnamefont {D.}~\bibnamefont {Nuzhnyy}},
  \bibinfo {author} {\bibfnamefont {S.}~\bibnamefont {Kamba}}, \bibinfo
  {author} {\bibfnamefont {D.~A.}\ \bibnamefont {Muller}}, \bibinfo {author}
  {\bibfnamefont {I.}~\bibnamefont {Takeuchi}}, \bibinfo {author}
  {\bibfnamefont {J.~C.}\ \bibnamefont {Booth}}, \bibinfo {author}
  {\bibfnamefont {C.~J.}\ \bibnamefont {Fennie}}, \ and\ \bibinfo {author}
  {\bibfnamefont {D.~G.}\ \bibnamefont {Schlom}},\ }\href {\doibase
  10.1038/nature12582} {\bibfield  {journal} {\bibinfo  {journal} {Nature}\
  }\textbf {\bibinfo {volume} {502}},\ \bibinfo {pages} {532} (\bibinfo {year}
  {2013})}\BibitemShut {NoStop}%
\bibitem [{\citenamefont {Nozaki}\ \emph {et~al.}(1991)\citenamefont {Nozaki},
  \citenamefont {Yoshikawa}, \citenamefont {Wada}, \citenamefont {Yamauchi},\
  and\ \citenamefont {Tanaka}}]{Nozaki1991}%
  \BibitemOpen
  \bibfield  {author} {\bibinfo {author} {\bibfnamefont {A.}~\bibnamefont
  {Nozaki}}, \bibinfo {author} {\bibfnamefont {H.}~\bibnamefont {Yoshikawa}},
  \bibinfo {author} {\bibfnamefont {T.}~\bibnamefont {Wada}}, \bibinfo {author}
  {\bibfnamefont {H.}~\bibnamefont {Yamauchi}}, \ and\ \bibinfo {author}
  {\bibfnamefont {S.}~\bibnamefont {Tanaka}},\ }\href {\doibase
  10.1103/PhysRevB.43.181} {\bibfield  {journal} {\bibinfo  {journal} {Physical
  Review B}\ }\textbf {\bibinfo {volume} {43}},\ \bibinfo {pages} {181}
  (\bibinfo {year} {1991})}\BibitemShut {NoStop}%
\bibitem [{\citenamefont {Sugiyama}\ \emph {et~al.}(2014)\citenamefont
  {Sugiyama}, \citenamefont {Nozaki}, \citenamefont {Umegaki}, \citenamefont
  {Higemoto}, \citenamefont {Ansaldo}, \citenamefont {Brewer}, \citenamefont
  {Sakurai}, \citenamefont {Kao}, \citenamefont {Yang},\ and\ \citenamefont
  {M{\aa}nsson}}]{Sugiyama2014}%
  \BibitemOpen
  \bibfield  {author} {\bibinfo {author} {\bibfnamefont {J.}~\bibnamefont
  {Sugiyama}}, \bibinfo {author} {\bibfnamefont {H.}~\bibnamefont {Nozaki}},
  \bibinfo {author} {\bibfnamefont {I.}~\bibnamefont {Umegaki}}, \bibinfo
  {author} {\bibfnamefont {W.}~\bibnamefont {Higemoto}}, \bibinfo {author}
  {\bibfnamefont {E.~J.}\ \bibnamefont {Ansaldo}}, \bibinfo {author}
  {\bibfnamefont {J.~H.}\ \bibnamefont {Brewer}}, \bibinfo {author}
  {\bibfnamefont {H.}~\bibnamefont {Sakurai}}, \bibinfo {author} {\bibfnamefont
  {T.-H.}\ \bibnamefont {Kao}}, \bibinfo {author} {\bibfnamefont {H.-D.}\
  \bibnamefont {Yang}}, \ and\ \bibinfo {author} {\bibfnamefont
  {M.}~\bibnamefont {M{\aa}nsson}},\ }\href {\doibase
  10.1103/PhysRevB.89.020402} {\bibfield  {journal} {\bibinfo  {journal}
  {Physical Review B}\ }\textbf {\bibinfo {volume} {89}},\ \bibinfo {pages}
  {020402(R)} (\bibinfo {year} {2014})}\BibitemShut {NoStop}%
\bibitem [{\citenamefont {Sakurai}(2015)}]{Sakurai2015}%
  \BibitemOpen
  \bibfield  {author} {\bibinfo {author} {\bibfnamefont {H.}~\bibnamefont
  {Sakurai}},\ }\href {\doibase 10.1016/j.phpro.2015.12.107} {\bibfield
  {journal} {\bibinfo  {journal} {Physics Procedia}\ }\textbf {\bibinfo
  {volume} {75}},\ \bibinfo {pages} {829} (\bibinfo {year} {2015})}\BibitemShut
  {NoStop}%
\bibitem [{\citenamefont {Chan}\ \emph {et~al.}(2000)\citenamefont {Chan},
  \citenamefont {Levin}, \citenamefont {Vanderah}, \citenamefont {Geyer},\ and\
  \citenamefont {Roth}}]{Chan2000}%
  \BibitemOpen
  \bibfield  {author} {\bibinfo {author} {\bibfnamefont {J.~Y.}\ \bibnamefont
  {Chan}}, \bibinfo {author} {\bibfnamefont {I.}~\bibnamefont {Levin}},
  \bibinfo {author} {\bibfnamefont {T.}~\bibnamefont {Vanderah}}, \bibinfo
  {author} {\bibfnamefont {R.}~\bibnamefont {Geyer}}, \ and\ \bibinfo {author}
  {\bibfnamefont {R.}~\bibnamefont {Roth}},\ }\href {\doibase
  10.1016/S1466-6049(00)00014-3} {\bibfield  {journal} {\bibinfo  {journal}
  {International Journal of Inorganic Materials}\ }\textbf {\bibinfo {volume}
  {2}},\ \bibinfo {pages} {107} (\bibinfo {year} {2000})}\BibitemShut {NoStop}%
\bibitem [{\citenamefont {Haislmaier}\ \emph {et~al.}(2016)\citenamefont
  {Haislmaier}, \citenamefont {Stone}, \citenamefont {Alem},\ and\
  \citenamefont {Engel-Herbert}}]{Haislmaier2016}%
  \BibitemOpen
  \bibfield  {author} {\bibinfo {author} {\bibfnamefont {R.~C.}\ \bibnamefont
  {Haislmaier}}, \bibinfo {author} {\bibfnamefont {G.}~\bibnamefont {Stone}},
  \bibinfo {author} {\bibfnamefont {N.}~\bibnamefont {Alem}}, \ and\ \bibinfo
  {author} {\bibfnamefont {R.}~\bibnamefont {Engel-Herbert}},\ }\href {\doibase
  10.1063/1.4959180} {\bibfield  {journal} {\bibinfo  {journal} {Applied
  Physics Letters}\ }\textbf {\bibinfo {volume} {109}},\ \bibinfo {pages}
  {043102} (\bibinfo {year} {2016})}\BibitemShut {NoStop}%
\bibitem [{\citenamefont {Haeni}\ \emph {et~al.}(2001)\citenamefont {Haeni},
  \citenamefont {Theis}, \citenamefont {Schlom}, \citenamefont {Tian},
  \citenamefont {Pan}, \citenamefont {Chang}, \citenamefont {Takeuchi},\ and\
  \citenamefont {Xiang}}]{Haeni2001}%
  \BibitemOpen
  \bibfield  {author} {\bibinfo {author} {\bibfnamefont {J.~H.}\ \bibnamefont
  {Haeni}}, \bibinfo {author} {\bibfnamefont {C.~D.}\ \bibnamefont {Theis}},
  \bibinfo {author} {\bibfnamefont {D.~G.}\ \bibnamefont {Schlom}}, \bibinfo
  {author} {\bibfnamefont {W.}~\bibnamefont {Tian}}, \bibinfo {author}
  {\bibfnamefont {X.~Q.}\ \bibnamefont {Pan}}, \bibinfo {author} {\bibfnamefont
  {H.}~\bibnamefont {Chang}}, \bibinfo {author} {\bibfnamefont
  {I.}~\bibnamefont {Takeuchi}}, \ and\ \bibinfo {author} {\bibfnamefont
  {X.-D.}\ \bibnamefont {Xiang}},\ }\href {\doibase 10.1063/1.1371788}
  {\bibfield  {journal} {\bibinfo  {journal} {Applied Physics Letters}\
  }\textbf {\bibinfo {volume} {78}},\ \bibinfo {pages} {3292} (\bibinfo {year}
  {2001})}\BibitemShut {NoStop}%
\bibitem [{\citenamefont {Isawa}\ and\ \citenamefont
  {Nagano}(2001)}]{Isawa2001}%
  \BibitemOpen
  \bibfield  {author} {\bibinfo {author} {\bibfnamefont {K.}~\bibnamefont
  {Isawa}}\ and\ \bibinfo {author} {\bibfnamefont {M.}~\bibnamefont {Nagano}},\
  }\href {\doibase 10.1016/S0921-4534(01)00245-3} {\bibfield  {journal}
  {\bibinfo  {journal} {Physica C: Superconductivity}\ }\textbf {\bibinfo
  {volume} {357-360}},\ \bibinfo {pages} {359} (\bibinfo {year}
  {2001})}\BibitemShut {NoStop}%
\bibitem [{\citenamefont {Wu}\ \emph {et~al.}(2010)\citenamefont {Wu},
  \citenamefont {Agrawal}, \citenamefont {Becerril}, \citenamefont {Bao},
  \citenamefont {Liu}, \citenamefont {Chen},\ and\ \citenamefont
  {Peumans}}]{Wu2010}%
  \BibitemOpen
  \bibfield  {author} {\bibinfo {author} {\bibfnamefont {J.}~\bibnamefont
  {Wu}}, \bibinfo {author} {\bibfnamefont {M.}~\bibnamefont {Agrawal}},
  \bibinfo {author} {\bibfnamefont {H.~A.}\ \bibnamefont {Becerril}}, \bibinfo
  {author} {\bibfnamefont {Z.}~\bibnamefont {Bao}}, \bibinfo {author}
  {\bibfnamefont {Z.}~\bibnamefont {Liu}}, \bibinfo {author} {\bibfnamefont
  {Y.}~\bibnamefont {Chen}}, \ and\ \bibinfo {author} {\bibfnamefont
  {P.}~\bibnamefont {Peumans}},\ }\href {\doibase 10.1021/nn900728d} {\bibfield
   {journal} {\bibinfo  {journal} {ACS Nano}\ }\textbf {\bibinfo {volume}
  {4}},\ \bibinfo {pages} {43} (\bibinfo {year} {2010})}\BibitemShut {NoStop}%
\bibitem [{\citenamefont {Macquart}\ \emph
  {et~al.}(2010{\natexlab{b}})\citenamefont {Macquart}, \citenamefont
  {Kennedy},\ and\ \citenamefont {Avdeev}}]{Macquart2010_SrMoO3}%
  \BibitemOpen
  \bibfield  {author} {\bibinfo {author} {\bibfnamefont {R.~B.}\ \bibnamefont
  {Macquart}}, \bibinfo {author} {\bibfnamefont {B.~J.}\ \bibnamefont
  {Kennedy}}, \ and\ \bibinfo {author} {\bibfnamefont {M.}~\bibnamefont
  {Avdeev}},\ }\href {\doibase 10.1016/j.jssc.2009.11.005} {\bibfield
  {journal} {\bibinfo  {journal} {Journal of Solid State Chemistry}\ }\textbf
  {\bibinfo {volume} {183}},\ \bibinfo {pages} {250} (\bibinfo {year}
  {2010}{\natexlab{b}})}\BibitemShut {NoStop}%
\bibitem [{\citenamefont {Mizoguchi}\ \emph {et~al.}(2000)\citenamefont
  {Mizoguchi}, \citenamefont {Kitamura}, \citenamefont {Fukumi}, \citenamefont
  {Mihara}, \citenamefont {Nishii}, \citenamefont {Nakamura}, \citenamefont
  {Kikuchi}, \citenamefont {Hosono},\ and\ \citenamefont
  {Kawazoe}}]{Mizoguchi2000}%
  \BibitemOpen
  \bibfield  {author} {\bibinfo {author} {\bibfnamefont {H.}~\bibnamefont
  {Mizoguchi}}, \bibinfo {author} {\bibfnamefont {N.}~\bibnamefont {Kitamura}},
  \bibinfo {author} {\bibfnamefont {K.}~\bibnamefont {Fukumi}}, \bibinfo
  {author} {\bibfnamefont {T.}~\bibnamefont {Mihara}}, \bibinfo {author}
  {\bibfnamefont {J.}~\bibnamefont {Nishii}}, \bibinfo {author} {\bibfnamefont
  {M.}~\bibnamefont {Nakamura}}, \bibinfo {author} {\bibfnamefont
  {N.}~\bibnamefont {Kikuchi}}, \bibinfo {author} {\bibfnamefont
  {H.}~\bibnamefont {Hosono}}, \ and\ \bibinfo {author} {\bibfnamefont
  {H.}~\bibnamefont {Kawazoe}},\ }\href {\doibase 10.1063/1.373111} {\bibfield
  {journal} {\bibinfo  {journal} {Journal of Applied Physics}\ }\textbf
  {\bibinfo {volume} {87}},\ \bibinfo {pages} {4617} (\bibinfo {year}
  {2000})}\BibitemShut {NoStop}%
\bibitem [{\citenamefont {Nagai}\ \emph {et~al.}(2005)\citenamefont {Nagai},
  \citenamefont {Shirakawa}, \citenamefont {Ikeda}, \citenamefont {Iwasaki},
  \citenamefont {Nishimura},\ and\ \citenamefont {Kosaka}}]{Nagai2005}%
  \BibitemOpen
  \bibfield  {author} {\bibinfo {author} {\bibfnamefont {I.}~\bibnamefont
  {Nagai}}, \bibinfo {author} {\bibfnamefont {N.}~\bibnamefont {Shirakawa}},
  \bibinfo {author} {\bibfnamefont {S.~I.}\ \bibnamefont {Ikeda}}, \bibinfo
  {author} {\bibfnamefont {R.}~\bibnamefont {Iwasaki}}, \bibinfo {author}
  {\bibfnamefont {H.}~\bibnamefont {Nishimura}}, \ and\ \bibinfo {author}
  {\bibfnamefont {M.}~\bibnamefont {Kosaka}},\ }\href {\doibase
  10.1063/1.1992671} {\bibfield  {journal} {\bibinfo  {journal} {Applied
  Physics Letters}\ }\textbf {\bibinfo {volume} {87}},\ \bibinfo {pages} {2}
  (\bibinfo {year} {2005})}\BibitemShut {NoStop}%
\bibitem [{\citenamefont {Haule}\ and\ \citenamefont
  {Kotliar}(2009)}]{Haule2009}%
  \BibitemOpen
  \bibfield  {author} {\bibinfo {author} {\bibfnamefont {K.}~\bibnamefont
  {Haule}}\ and\ \bibinfo {author} {\bibfnamefont {G.}~\bibnamefont
  {Kotliar}},\ }\href {\doibase 10.1088/1367-2630/11/2/025021} {\bibfield
  {journal} {\bibinfo  {journal} {New Journal of Physics}\ }\textbf {\bibinfo
  {volume} {11}},\ \bibinfo {pages} {025021} (\bibinfo {year}
  {2009})}\BibitemShut {NoStop}%
\bibitem [{\citenamefont {Deng}\ \emph {et~al.}(2019)\citenamefont {Deng},
  \citenamefont {Stadler}, \citenamefont {Haule}, \citenamefont {Weichselbaum},
  \citenamefont {von Delft},\ and\ \citenamefont {Kotliar}}]{Deng2017}%
  \BibitemOpen
  \bibfield  {author} {\bibinfo {author} {\bibfnamefont {X.}~\bibnamefont
  {Deng}}, \bibinfo {author} {\bibfnamefont {K.~M.}\ \bibnamefont {Stadler}},
  \bibinfo {author} {\bibfnamefont {K.}~\bibnamefont {Haule}}, \bibinfo
  {author} {\bibfnamefont {A.}~\bibnamefont {Weichselbaum}}, \bibinfo {author}
  {\bibfnamefont {J.}~\bibnamefont {von Delft}}, \ and\ \bibinfo {author}
  {\bibfnamefont {G.}~\bibnamefont {Kotliar}},\ }\href {\doibase
  10.1038/s41467-019-10257-2} {\bibfield  {journal} {\bibinfo  {journal}
  {Nature Communications}\ }\textbf {\bibinfo {volume} {10}},\ \bibinfo {pages}
  {2721} (\bibinfo {year} {2019})}\BibitemShut {NoStop}%
\bibitem [{\citenamefont {de' Medici}\ \emph {et~al.}(2011)\citenamefont {de'
  Medici}, \citenamefont {Mravlje},\ and\ \citenamefont
  {Georges}}]{DeMedici2011}%
  \BibitemOpen
  \bibfield  {author} {\bibinfo {author} {\bibfnamefont {L.}~\bibnamefont {de'
  Medici}}, \bibinfo {author} {\bibfnamefont {J.}~\bibnamefont {Mravlje}}, \
  and\ \bibinfo {author} {\bibfnamefont {A.}~\bibnamefont {Georges}},\ }\href
  {\doibase 10.1103/PhysRevLett.107.256401} {\bibfield  {journal} {\bibinfo
  {journal} {Physical Review Letters}\ }\textbf {\bibinfo {volume} {107}},\
  \bibinfo {pages} {256401} (\bibinfo {year} {2011})}\BibitemShut {NoStop}%
\bibitem [{\citenamefont {Wang}\ \emph {et~al.}(2018)\citenamefont {Wang},
  \citenamefont {Prakash}, \citenamefont {Dong}, \citenamefont {Truttmann},
  \citenamefont {Bucsek}, \citenamefont {James}, \citenamefont {Fong},
  \citenamefont {Kim}, \citenamefont {Ryan}, \citenamefont {Zhou},
  \citenamefont {Birol},\ and\ \citenamefont {Jalan}}]{Wang2018}%
  \BibitemOpen
  \bibfield  {author} {\bibinfo {author} {\bibfnamefont {T.}~\bibnamefont
  {Wang}}, \bibinfo {author} {\bibfnamefont {A.}~\bibnamefont {Prakash}},
  \bibinfo {author} {\bibfnamefont {Y.}~\bibnamefont {Dong}}, \bibinfo {author}
  {\bibfnamefont {T.}~\bibnamefont {Truttmann}}, \bibinfo {author}
  {\bibfnamefont {A.}~\bibnamefont {Bucsek}}, \bibinfo {author} {\bibfnamefont
  {R.}~\bibnamefont {James}}, \bibinfo {author} {\bibfnamefont {D.~D.}\
  \bibnamefont {Fong}}, \bibinfo {author} {\bibfnamefont {J.-W.}\ \bibnamefont
  {Kim}}, \bibinfo {author} {\bibfnamefont {P.~J.}\ \bibnamefont {Ryan}},
  \bibinfo {author} {\bibfnamefont {H.}~\bibnamefont {Zhou}}, \bibinfo {author}
  {\bibfnamefont {T.}~\bibnamefont {Birol}}, \ and\ \bibinfo {author}
  {\bibfnamefont {B.}~\bibnamefont {Jalan}},\ }\href {\doibase
  10.1021/acsami.8b16592} {\bibfield  {journal} {\bibinfo  {journal} {ACS
  Applied Materials {\&} Interfaces}\ }\textbf {\bibinfo {volume} {10}},\
  \bibinfo {pages} {43802} (\bibinfo {year} {2018})}\BibitemShut {NoStop}%
\bibitem [{\citenamefont {Eklund}\ \emph {et~al.}(2009)\citenamefont {Eklund},
  \citenamefont {Fennie},\ and\ \citenamefont {Rabe}}]{Eklund2009}%
  \BibitemOpen
  \bibfield  {author} {\bibinfo {author} {\bibfnamefont {C.-J.}\ \bibnamefont
  {Eklund}}, \bibinfo {author} {\bibfnamefont {C.~J.}\ \bibnamefont {Fennie}},
  \ and\ \bibinfo {author} {\bibfnamefont {K.~M.}\ \bibnamefont {Rabe}},\
  }\href {\doibase 10.1103/PhysRevB.79.220101} {\bibfield  {journal} {\bibinfo
  {journal} {Physical Review B}\ }\textbf {\bibinfo {volume} {79}},\ \bibinfo
  {pages} {220101(R)} (\bibinfo {year} {2009})}\BibitemShut {NoStop}%
\end{thebibliography}
\end{document}